\pdfoutput=1
\documentclass[conference]{IEEEtran}

\makeatletter

\def\ps@IEEEtitlepagestyle{%
  \def\@oddfoot{\mycopyrightnotice}%
  \def\@evenfoot{}%
}

\usepackage{blindtext}
\usepackage{eso-pic}
\IEEEoverridecommandlockouts
\usepackage{cite}
\usepackage{amsmath,amssymb,amsfonts}
\usepackage{algorithmic}
\usepackage{graphicx}
\usepackage{textcomp}
\usepackage{xcolor}
\def\BibTeX{{\rm B\kern-.05em{\sc i\kern-.025em b}\kern-.08em
    T\kern-.1667em\lower.7ex\hbox{E}\kern-.125emX}}
    
\usepackage{eso-pic}

\usepackage{multirow}
\usepackage{hyperref}
\usepackage{cleveref}
\usepackage{xcolor}
\usepackage{longtable}
\usepackage{siunitx}
\PassOptionsToPackage{hyphens}{url}
\usepackage{card}
\usepackage{microtype}
\usepackage{tabularx}
\usepackage{enumitem}
\usepackage{subcaption}
\usepackage{graphicx}
\usepackage{annotation}

\begin{document}
\title{\vspace*{1cm} Big Tech-Funded AI Papers Have Higher Citation Impact, Greater Insularity, and Larger Recency Bias\\
\thanks{This work was partially supported by the Lower
Saxony Ministry of Science and Culture and the
VW Foundation.}
}

\author{
\IEEEauthorblockN{Max Martin Gnewuch}
\IEEEauthorblockA{\textit{University of G{\"o}ttingen} \\
G{\"o}ttingen, Lower Saxony, Germany \\
maxmartin.gnewuch@stud.uni-goettingen.de} \\[0.2cm]
\IEEEauthorblockN{Terry Ruas}
\IEEEauthorblockA{\textit{University of G{\"o}ttingen} \\
G{\"o}ttingen, Lower Saxony, Germany \\
ruas@uni-goettingen.de} \and
\IEEEauthorblockN{Jan Philip Wahle}
\IEEEauthorblockA{\textit{University of G{\"o}ttingen} \\
G{\"o}ttingen, Lower Saxony, Germany \\
wahle@uni-goettingen.de}
\\[0.2cm]
\IEEEauthorblockN{Bela Gipp}
\IEEEauthorblockA{\textit{University of G{\"o}ttingen} \\
G{\"o}ttingen, Lower Saxony, Germany \\
gipp@uni-goettingen.de}
}

\maketitle

\AddAnnotationRef

\begin{abstract}
Over the past four decades, artificial intelligence (AI) research has flourished at the nexus of academia and industry.
However, Big Tech companies have increasingly acquired the edge in computational resources, big data, and talent.
So far, it has been largely unclear how many papers the industry funds, how their citation impact compares to non-funded papers, and what drives industry interest.
This study fills that gap and quantifies the number of industry-funded papers at \num{10} top AI conferences (e.g., ICLR, CVPR, AAAI, ACL) and their citation influence by analyzing $\approx$\num{49.8}\,K papers, $\approx$\num{1.8}\,M citations from AI papers to other papers, and $\approx$\num{2.3}\,M citations from other papers to AI papers from \num{1998}--\num{2022} in Scopus.
We investigate the volume and evolution of industry funding in AI research, the citation impact of the papers, the diversity and temporal range of their citations, and the subfields in which industry predominantly acts through \num{7} research questions.
Our findings reveal that the industry present has grown markedly since 2015, from less than \SI{2}{\percent} to more than \SI{11}{\percent} in 2020.
Between \num{2018} and \num{2022}, \SI{12}{\percent} of industry-funded papers achieved high citation rates as measured by h5-index, compared to \SI{4}{\percent} of non-industry-funded papers and \SI{2}{\percent} of non-funded papers. 
Top AI conferences engage more with industry-funded research than non-funded research, as measured by our newly proposed metric, the \textit{Citation Preference Ratio} ($CPR$).
We show that industry-funded research is increasingly insular --- citing predominantly other industry-funded papers while referencing fewer non-funded papers.
Furthermore, industry-funded work cites more recent work and fewer older papers than non-funded works.
These findings reveal new trends in AI research funding, including a shift towards (\num{1}) more industry-funded papers and their growing citation impact, (\num{2}) greater insularity of industry-funded works than non-funded works, and (\num{3}) preference of industry-funded research to cite recent work.
While industry funding contributes markedly to AI research, these new trends also raise questions about Big Tech's allocation of resources and potential control over research topics.
All data and code are publicly available: \url{https://github.com/Peerzival/impact-big-tech-funding}.
\end{abstract}

\begin{IEEEkeywords}
Scientometrics, Big Tech, Funding, Power Dynamics, Monopolization, Echo Chambers, Ethical AI, Bias.
\end{IEEEkeywords}

\section{Introduction}

\begin{figure}[tbp]
  \centering
  \includegraphics[width=\columnwidth]{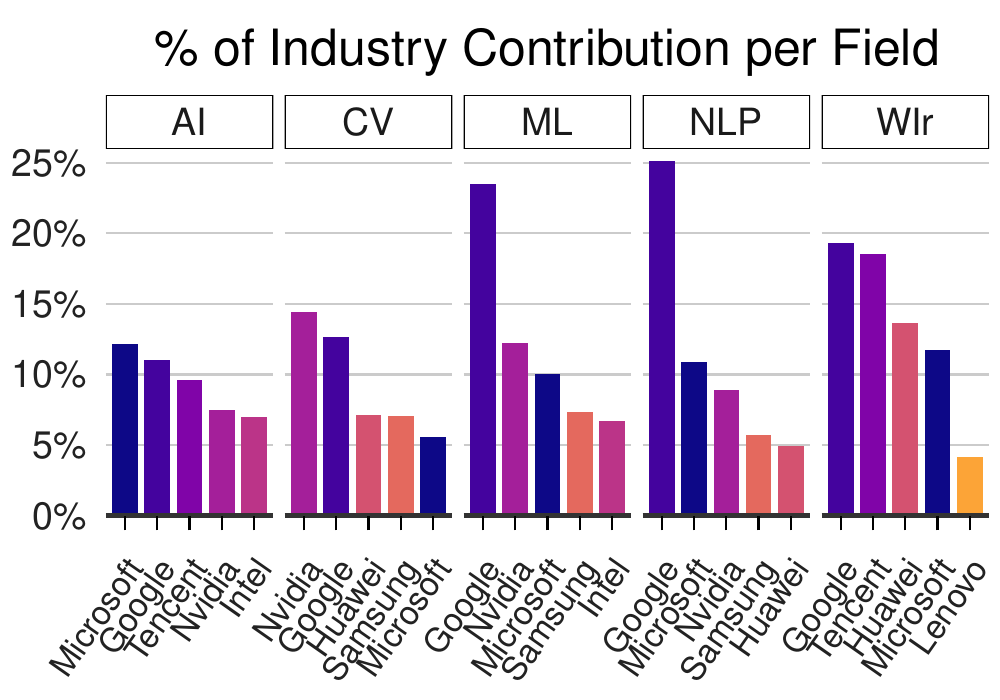}
  \caption{The share of the top \num{5} companies regarding all industry funding across key domains between \num{2018} to \num{2022}.} %
  \vspace*{-.45cm}
  \label{fig:topFundingAgencies}
\end{figure}

Artificial Intelligence (AI) has evolved into a ``general purpose technology''~\cite{ahmed2020dedemocratization, verdegem2024dismantling} comparable to other historic economic transformations such as the steam engine, electrification, and the Internet~\cite{verdegem2024dismantling}, offering new opportunities across industries \cite{NBERw24449}.
This evolution can reshape society by affecting job markets~\cite{frey2017future}, healthcare~\cite{jiang2017artificial}, and addressing global challenges like climate change~\cite{dobbe2019ai}.
However, integrating AI applications can lead to unexpected consequences and introduce new types of risks.  
For example, developing advanced weapons and autonomous military systems could trigger an arms race, escalating global tensions and undermining security \cite{brundage2018malicious}. 
If policymakers do not support displaced workers adequately, this trend may lead to economic instability \cite{frey2017future}.
Recognizing these risks alongside opportunities, the development of AI systems must be inclusive, ensuring that their benefits are widely accessible rather than limited to a privileged few \cite{ahmed2020dedemocratization,plan2016national,koenecke2020racial}. 
Research teams that are diverse in terms of age, gender, race, and ethnicity and that engage with diverse literature across time and fields are better equipped to create more democratic, fair, and unbiased AI \cite{kuhlman2020computation, west2019discriminating}.
However, early evidence suggests that AI development is becoming increasingly insular~\cite{klinger2022narrowingairesearch} with few powerful actors leading innovations \cite{ahmed2020dedemocratization,whittaker2021steep,ahmed2023growing}. 

\noindent \textbf{The role of Big Tech.}
The industrial landscape of AI experiences strong interest by a small number of large companies, mainly in the US and China (\Cref{fig:topFundingAgencies}), often referred to as \textit{Big Tech} (e.g., Meta, Google, Amazon, Nvidia, Microsoft, Baidu, Tencent) \cite{montes2019distributed, abdalla-etal-2023-elephant, ahmed2020dedemocratization, riedl2020ai, verdegem2024dismantling}. 
These companies have access to \num{3} key resources essential for modern AI research: big data, computational power, and access to a highly skilled AI workforce \cite{montes2019distributed, riedl2020ai, ahmed2023growing, verdegem2024dismantling}. 
This oligopolistic/monopolistic control over critical resources could grant a disproportionate amount of power to a small number of corporations \cite{montes2019distributed, ahmed2023growing}, who contribute to shaping what is examined (and not examined) in AI research \cite{whittaker2021steep}.
As a result, there is growing concern about scientific independence, influence, and power concentration in AI research \cite{abdalla-etal-2021-tobacco-tech, whittaker2021steep, ahmed2023growing}.

\noindent \textbf{Tensions between commercial and public interests.}
The goals of industry and those of the public or society can differ markedly.
Connections between industry and the AI research community mean that the questions and incentives shaping the field are not always within the control of individual researchers \cite{whittaker2021steep}. 
The tech industry's incentives can influence the trajectory of a field, including which questions are deemed worth pursuing and which answers lead to grants, awards, and tenure \cite{whittaker2021steep}.
The increasing industry investment in AI research does not diminish the potential for substantial societal benefits. 
However, commercial motivations often drive companies to prioritize profit-oriented goals and topics. 
Such incentives can but do not have to align with the public interest and result in beneficial tools, hardware, or software. 
Without public alternatives, AI may follow a similar pattern observed in the pharmaceutical industry, where investments can overlook the needs of those that deserve particular attention, such as lower-income groups \cite{trouiller2002drug, ahmed2023growing}.

\noindent \textbf{Research Gap.}
Prior work has examined Big Tech presence, including analyses of Big Tech funding for AI researchers \cite{abdalla-etal-2021-tobacco-tech}, the rising share of Big Tech–affiliated publications at top conferences \cite{ahmed2020dedemocratization}, the migration of researchers from academia to private sector and talent shifts, research focus, value differences, and impact disparities between academia and industry \cite{jurowetzki2021privatizationairesearcherscauses, klinger2022narrowingairesearch, birhane2022valuesencodedmachinelearning, gizinski2024big}.
These works rely on author affiliations instead of direct funding attribution and use a subset of publicly traded companies.
Further prior work lacks reference from Big Tech funding to other funding types, such as public funding and papers without funding.
By doing so, we aim to foster further discourse on the interplay between academic and industry research and advance a more nuanced understanding of their relationship.

\noindent \textbf{This study.}
In this work, we systematically assess the influence of funded papers by Big Tech and public funding types on the citation practices of AI papers.
Using the Scopus database, we compiled a new dataset of metadata associated with $\approx$\num{49.8}\,K AI papers published at \num{10} top conferences\footnote{According to \href{https://csrankings.org/\#/index?ai&vision&mlmining&nlp&inforet&world}{csranking.org}} from \num{1998} to \num{2022}, along with $\approx$\num{1.8}\,M citations from AI papers and $\approx$\num{2.3}\,M citations to AI papers. 
The dataset contains both ties of authors to industry through author affiliations and extracted funding statements through acknowledgement sections of papers. 
In this analysis, we mainly focus on the acknowledgement sections of direct funding to a paper as they are a high-precision way of tracing direct funding.
In our dataset, each citation includes details about the year of publication, the funding agency identified in the paper for both industry and non-industry funders, and the field of study for both the papers \textit{cited} by AI research and those \textit{citing} AI research. 
We use this new dataset to address different questions of industry in AI:

\begin{enumerate}[leftmargin=*,listparindent=0em]
    \item \textbf{Evolution of industry presence in AI}: How has the volume of industry-funded research evolved over time? Which companies have high publication output? What specific fields within AI attract the greatest attention from industry?
    
    \item \textbf{Engagement of the AI community}: How much do papers from top AI venues cite industry-funded research? How has this citation behaviour changed over time? How does the citation impact of industry-funded papers compare to other AI papers? Which companies have the highest citation impact?
    
    \item \textbf{Insularity of industry-funded research}: Does industry-funded research predominantly cite other industry-funded research instead of research from other funding sources? How much does industry-funded research cite different academic fields compared to other AI papers?
    
    \item \textbf{Recency bias of industry-funded research}: How far back in time do industry-funded papers cite on average compared to other AI papers? Do industry-funded papers cite more recent or older work than other AI papers?
\end{enumerate}

We show that the industry present has grown markedly since 2015 from less than \SI{2}{\percent} to more than \SI{11}{\percent} in 2020.
Since 2020, the percentage of industry-funded papers has slightly declined by \SI{14}{\percent}, while the AI community's engagement in industry-funded research has increased. 
\SI{12}{\percent} of industry-funded papers have a high citation impact as measured by h5-index, compared to \SI{4}{\percent} of non-industry-funded and \SI{2}{\percent} of non-funded papers.
We provide evidence that industry-funded research is increasingly insular, citing other industry-funded research \SI{2}{\percent} more often than other funding types.
Furthermore, our experiments show that industry-funded research references more recent papers and cites fewer older ones than non-funded research.

These results have potential implications for the academic community. 
The decreasing presence of industry-funded papers could signal Big Tech increasingly using their own channels for publication while moving out of these venues, which also implies reduced funding for venues --- a key funding source that many venues still rely on.
Our results suggest that publishing in top AI conferences has become less vital for Big Tech to achieve high citation rates and visibility for industry-funded research. 
The disproportionate engagement with research that Big Tech is funding and the limited diversity of companies~\cite{west2019discriminating,kuhlman2020computation} may affect long-term innovation.
Our results suggest that a few AI companies are increasingly impacting modern AI research as measured by citations.
Evidence exists that these companies may not reflect the interests of the broader population~\cite{west2019discriminating,kuhlman2020computation}, and our results support that papers from Big Tech prioritize recent literature disproportionately and have insular citation cycles compared to other AI papers.

\section{Related Work}

\noindent\textbf{Big tech influence on AI research.}
Several studies have attempted to quantify Big Tech’s influence in AI research using various methods and data sources.
\cite{abdalla-etal-2021-tobacco-tech} revealed that \SI{59}{\percent} of papers published in top AI journals addressing ethical and societal implications feature at least one author with financial ties to Big Tech.
Similarly, \cite{ahmed2020dedemocratization} analyzed participation trends in AI conferences after the rise of deep learning in \num{2012}, finding a marked increase in representation from major technology companies.

\cite{jurowetzki2021privatizationairesearcherscauses} used bibliometric data to trace researcher migration from academia to industry. 
Their findings indicate that \SI{25}{\percent} of AI researchers at top \num{5} Nature Index institutions transition to industry, highlighting Big Tech's competitive offers to attract top talent.

The study settings of \cite{farber2024analyzing} allowed the authors to examine the link between private-sector affiliations and research impact, measured through citations and attention scores. Their analysis shows that the private sector is dominant in shaping AI research.

Work by \cite{gizinski2024big} compared citation networks and memetic propagation between Big Tech and academia. Their results suggest that Big Tech-affiliated papers are disproportionately cited, with the most impactful research stemming from collaborations between academia and industry.

Several studies have tried to analyze the content of AI papers and study the influence of Big Tech.
\cite{klinger2022narrowingairesearch} study subject co-occurrence in arXiv papers, while \cite{birhane2022valuesencodedmachinelearning} investigated funding sources, reporting an increasing presence of Big Tech.

Yet, there is a key gap in our understanding of Big Tech’s influence.
Most studies described above equate influence with institutional affiliation shares (with the exception of \cite{gizinski2024big} and \cite{farber2024analyzing}).
However, this approach does not consider the number of the paper’s citations, which is a key proxy for the impact of a paper \cite{singh2023forgotten,gizinski2024big,rungta2022geographic,mohammad-2020-examining}. 
Furthermore, the reliance on publicly traded companies excludes key non-profit research organizations (e.g., OpenAI), particularly in the fast-moving field of AI, where startups emerge fast and have a marked impact.

\noindent\textbf{Citation patterns in scientific works.}
Related work has attracted significant attention to citation patterns exploring various aspects such as self-citation \cite{della2008multi}, publication venue \cite{callaham2002journal, wahle2022d3}, paper quality \cite{buela2010analysis}, publication language \cite{lira2013influence}, geographic location \cite{rungta2022geographic}, gender \cite{mohammad-2020-gender,chatterjee2021gender,abdalla2023ethnicity}, and field of study \cite{costas2009scaling}.

Interdisciplinary research field engagement represents a core component of responsible research \cite{porter2008interdisciplinary,leydesdorff2019interdisciplinarity}.
Numerous breakthroughs have emerged from such interactions, including Einstein’s photoelectric effect \cite{einstein1905erzeugung} and Bohr’s atomic model \cite{bohr1913constitution}.
Similarly, medicine has significantly benefited from integrating neuroscience \cite{postel2012nature} and ecology \cite{pearce1989economics}.

An area of recent particular interest is the temporal aspect of citations.
\cite{verstak2014shouldersgiantsgrowingimpact} identified an increasing tendency to cite older papers between \num{1990} and \num{2013}.
\cite{singh2023forgotten} reported a recency bias in Natural Language Processing (NLP), showing that post-\num{2015} publications increasingly favour recent work.
\cite{wahle2024citation} analyzed citation trends in NLP and another academic field, highlighting the field's strong focus on recent research.
So far, it has been unclear whether differences exist in citation patterns in industry research and other AI work.
Understanding these patterns is crucial, given Big Tech's potential to shape AI research trajectories and their broader implications for society and science.

Our study addresses gaps from related work by exploring \num{7} novel research questions, including the volume of papers with acknowledgment of industry funding (Q1), citation behaviours related to funding types (Q2, Q3, Q4), the topics which influence-funded research cites (Q5), and how far back in time industry-funded papers cite (Q6, Q7).

\section{Methodology}
We derive a new dataset from Scopus\footnote{The in-house Scopus database maintained by the German Competence Center for Bibliometrics (Scopus-KB), \num{2024} 07 version.}, a database that includes papers published between \num{1902}--\num{2024}, totalling $\approx$\num{69.5}\,M papers, $\approx$\num{2.2}\,B citations, and providing funding information for $\approx$\num{21.0}\,M papers. 
Scopus extracts funding information primarily through the acknowledgment sections of papers, which we use as a lower bound. %
We provide an overview of the key dataset statistics in \Cref{tab:datasetStatistics}.
Scopus has a broad coverage of journals and conferences, high quality control and moderation standards, and many peer-reviewed publications with inclusive content coverage \cite{pranckutėScopus21}.

To trace citations from Big Tech to other papers, we construct a citation graph that extends \num{2} levels deep from the papers published at top AI conferences as part of our \textbf{data collection}.
We focus our analysis on the period from \num{2018} to \num{2022} to capture the marked impact of transformer-based models on AI research. 
We chose \num{2022} as the end date because some conference data only extend to \num{2022} in the Scopus version available. 
Next, in \textbf{data processing}, we determine whether the extracted names belong to Big Tech funders using manual and automatic methods.
The final dataset contains \num{49811} papers.

\subsection{Data Collection} \label{sec:DC}

For the analysis of AI subfields, we select \num{5} key domains: \textit{Artificial Intelligence} (AI), \textit{Computer Vision} (CV), \textit{Machine Learning} (ML), \textit{Natural Language Processing} (NLP), and \textit{Web \& Information Retrieval} (WIr), based on csranking.org\footnote{\url{https://csrankings.org/\#/index?ai&vision&mlmining&nlp&inforet&world}}. 
We include \num{2} top conferences per domain using the same rank and their h5-index. 
Of the \num{10} conferences, \num{7} are in the Scopus database, while the \num{3} unmatched conferences (i.e., ECCV, NeurIPS, and WWW) get replaced by those with the third-highest h5-index in their respective fields (i.e., ICCV, ICML, and WSDM). 
The final list of top AI conferences contains AAAI, IJCAI, CVPR, ICCV, ICML, ICLR, ACL, EMNLP, WSDM, and SIGIR. 
\Cref{tab:aiConferences} provides details of the selected top AI conferences.

    \begin{table}[t]
        \centering
        \small \begin{tabular}{lr}
            \toprule
            Time range               & 1902–2024 \\
            \midrule
            \#papers & \num{69491766} \\
            \#funded $\text{papers}^{\dag}$ & \num{21047938} \\
            \#citations & \num{2199264185} \\
            \#papers AI* & \num{114090} \\
            \#funded $\text{papers AI}^{\dag}$ & \num{45893} \\
            \#out-citations from AI & \num{3308618} \\ 
            \#in-citations to AI & \num{6012570} \\ 
            \bottomrule
            \addlinespace
            \multicolumn{2}{@{}l}{\footnotesize $^{\dag}\text{Lower bound.}$ *Sum of articles in \Cref{tab:aiConferences}.}
        \end{tabular}
        \caption{Overall dataset statistics.}
        
        \label{tab:datasetStatistics}
    \end{table}
    \begin{table}[t]
        \centering
        \small \begin{tabular}{lccc}
            \toprule
            \textbf{Field} & \textbf{Conference} & \textbf{Number of articles} & \textbf{h5-index} ($\downarrow$)\\
            \midrule
            \multirow{2}{*}{AI} & AAAI & \num{6793} & 212 \\
                                & IJCAI & \num{4203} & 133 \\
            \addlinespace
            \multirow{2}{*}{CV} & CVPR & \num{9923} & 422 \\
                                & ICCV* & \num{4035} & 291 \\
            \addlinespace
            \multirow{2}{*}{ML} & ICLR & \num{3778} & 303 \\
                                & ICML* & \num{5733} & 288 \\
            \addlinespace
            \multirow{2}{*}{NLP} & ACL & \num{8674} & 192 \\
                                & EMNLP & \num{5513} & 176 \\
            \addlinespace
            \multirow{2}{*}{WIr} & SIGIR & \num{1874} & 90 \\
                                & WSDM* & \num{707} & 77 \\
            \bottomrule
            \addlinespace
        \end{tabular}
        \caption{The selected top AI conferences ordered by field and increasing h5-index.}
        \label{tab:aiConferences}
    \end{table}

\noindent \textbf{Citation Graph.}
To define the search scope of Big Tech, we construct a citation graph.
The graph originates from papers published at the selected top conferences (root level). 
It expands through the outgoing citations of these papers (level~\num{1}) as well as papers cited by those in level~\num{1} (level~\num{2}).
We do the same for analogous incoming citations.
We extract the names of the funding agencies for each paper in the root papers and those in levels \num{1} and \num{2} and identify industry funders (IFs) contributing to AI research.
This two-level expansion addresses the limitations of previous approaches focusing only on Big Tech contributions in conferences, which extracted companies by market capitalization.
We allow papers outside of conferences because some key AI companies do not publish directly at AI conferences.
For example, organizations such as OpenAI, Hugging Face, and Mistral play a fundamental role in advancing the field, but they mostly publish through their content delivery networks.

\subsection{Data Processing}

The process of generating our list of IFs involved five distinct steps:

\begin{enumerate}[leftmargin=*,listparindent=0em]
    \item \textbf{Extraction of Funding Agencies}: 
    We extracted the names of funding agencies and the number of funded papers from Scopus.
    
    \item \textbf{Manual Analysis}: 
    For agencies with more than ten funded papers, we manually reviewed their names to determine their classification as IF.
    A funding agency was designated as IF if it was neither public nor non-profit.
    
    \item \textbf{Standardization of IF Names}: 
    We standardized IF names to address variations in company nomenclature and alternative descriptions for the same organization (e.g., Amazon, Amazon Research).
    
    \item \textbf{Automatic Analysis}: 
    We used fuzzy matching to examine the remaining funding agencies with fewer than ten occurrences. 
    A funding agency was classified as IF if its name included one of the \num{216} standardized IF names identified in the \num{3}. step. 
    
    \item \textbf{Integration of Results}: 
    Finally, we combined the results of the manual and automatic analyses into a unified list of IF names. 
    The full list of extracted funding agencies and number of funded papers can be found at \url{https://github.com/Peerzival/impact-big-tech-funding}.
\end{enumerate}

\noindent In total, we processed \num{78333} funding agencies and identified \num{3136} as IF.
\Cref{tab:fundingAgencyDatasetStatistics} summarizes key statistics for the dataset.

\begin{table}[tbp]
\small \centering    \caption{Overview of funding agency dataset statistics.}
    \label{tab:fundingAgencyDatasetStatistics}
        \begin{tabular}{lr}
            \toprule
            \textbf{Attribute}                & \textbf{Amount} \\
            \midrule
            Total funding agencies  & \num{78333} \\ 
            Funding agencies $\geq$ 10 occurrences  & \num{4206} \\
            Total corporate funding Agencies*  & \num{3136} \\
            Corporate funding agencies (manual analysis)  & \num{382} \\
            Corporate funding agencies after standardization  & \num{216} \\
            Corporate funding agencies (automatic analysis)  & \num{2754} \\
            \bottomrule
            \addlinespace
            \multicolumn{2}{@{}l}{\footnotesize *Sum of manual and automatic analysis.}
        \end{tabular}
    \vspace*{-.4cm}
\end{table}

\section{Experiments} \label{sec:experiments}

We use the dataset described above to answer a series of questions about industry funding in AI.

\noindent \textbf{Q1.}\label{sec:Q1}
\textit{How many papers have received industry funding in top AI conferences? How does this number vary by research field? Has this number stayed roughly the same, or has it changed markedly over the years?}

\noindent \textbf{Ans.}
We calculate the \textit{percentage of industry-funded papers} ($PIFP$) in a given year $y$ as: $PIFP(y) = \sum_{\forall f_i \in \digamma} \frac{ IF(f_i, y) }{ P(f_i, y) }$
where $IF(f_i, y)$ represents the number of industry-funded papers in field $f_i$ in year $y$, and $P(f_i, y)$ is the total number of papers in that field. 
$F$ is the set of all subfields in AI.
To isolate industry funding trends within individual subfields, we also compute the \textit{field-specific industry funding percentage }($FIFP$) for each year, where $FIFP(x, y) = \frac{ IF(x, y) }{ \sum_{\forall f_i \in \digamma} P(f_i, y) } \cdot 100$

\begin{figure*}[tbp]
    \centering
    \begin{subfigure}[ht]{.49\textwidth}
    \centering
        \includegraphics[scale = 0.35]{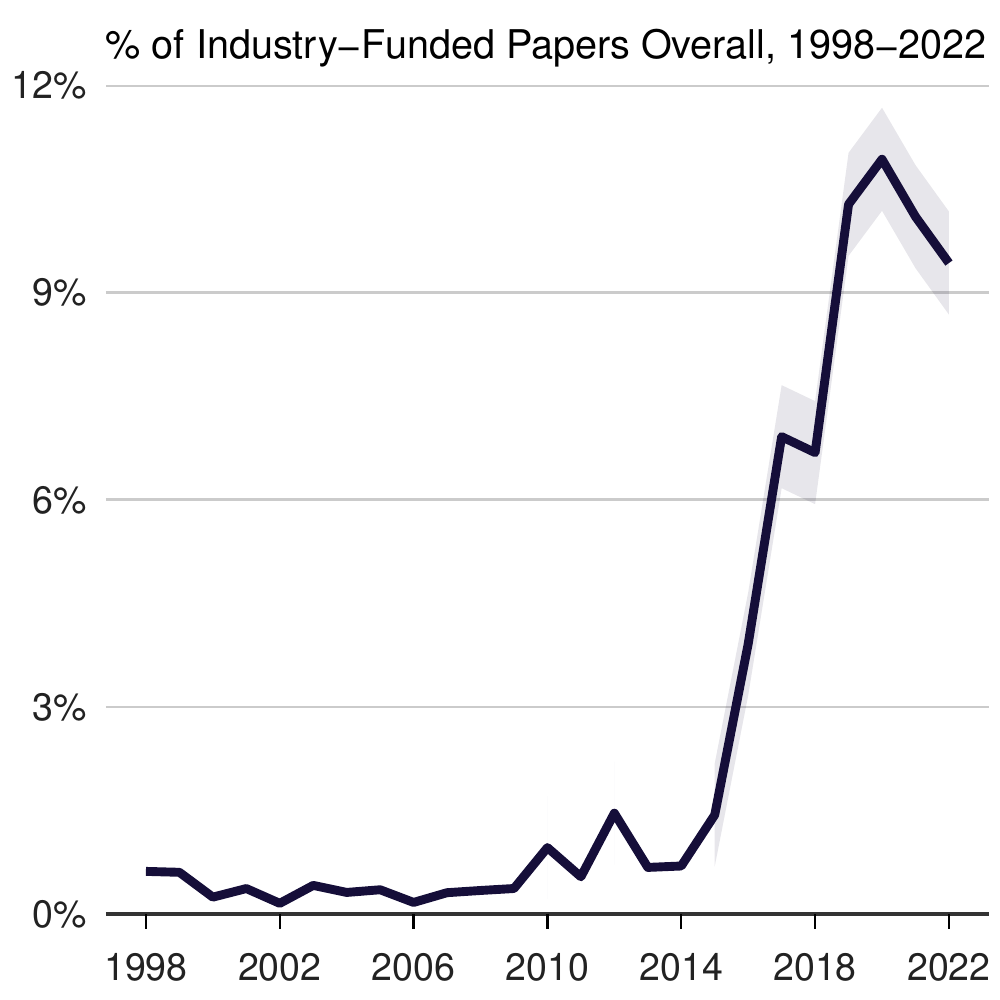}
        \caption{}
        \label{fig:subFundingPercentageOverall_98_23}
    \end{subfigure}
    \hfill
    \begin{subfigure}[ht]{.49\textwidth}
    \centering
        \includegraphics[scale = 0.35]{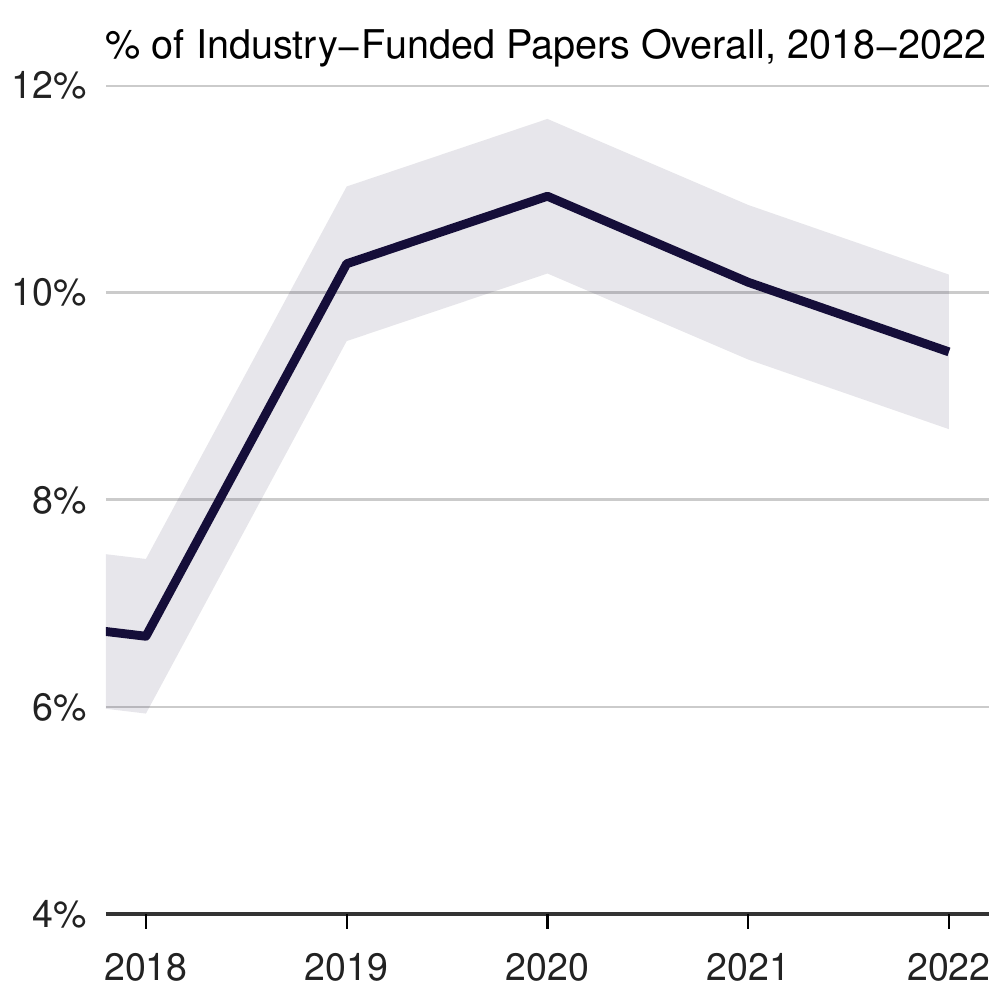}
        \caption{}
        \label{fig:subFundingPercentageOverall_18_23}
    \end{subfigure}
    \caption{The \textit{PIFP} (\subref{fig:subFundingPercentageOverall_98_23}) from \num{1998} to \num{2022} and (\subref{fig:subFundingPercentageOverall_18_23}) from \num{2018} to \num{2022}.}
    \label{fig:fundingPercentageYears}
    \vspace*{-.4cm}
\end{figure*}

\begin{figure*}[tbp]
    \centering
    \begin{subfigure}[ht]{.48\textwidth}
    \centering
        \includegraphics[scale = 0.36]{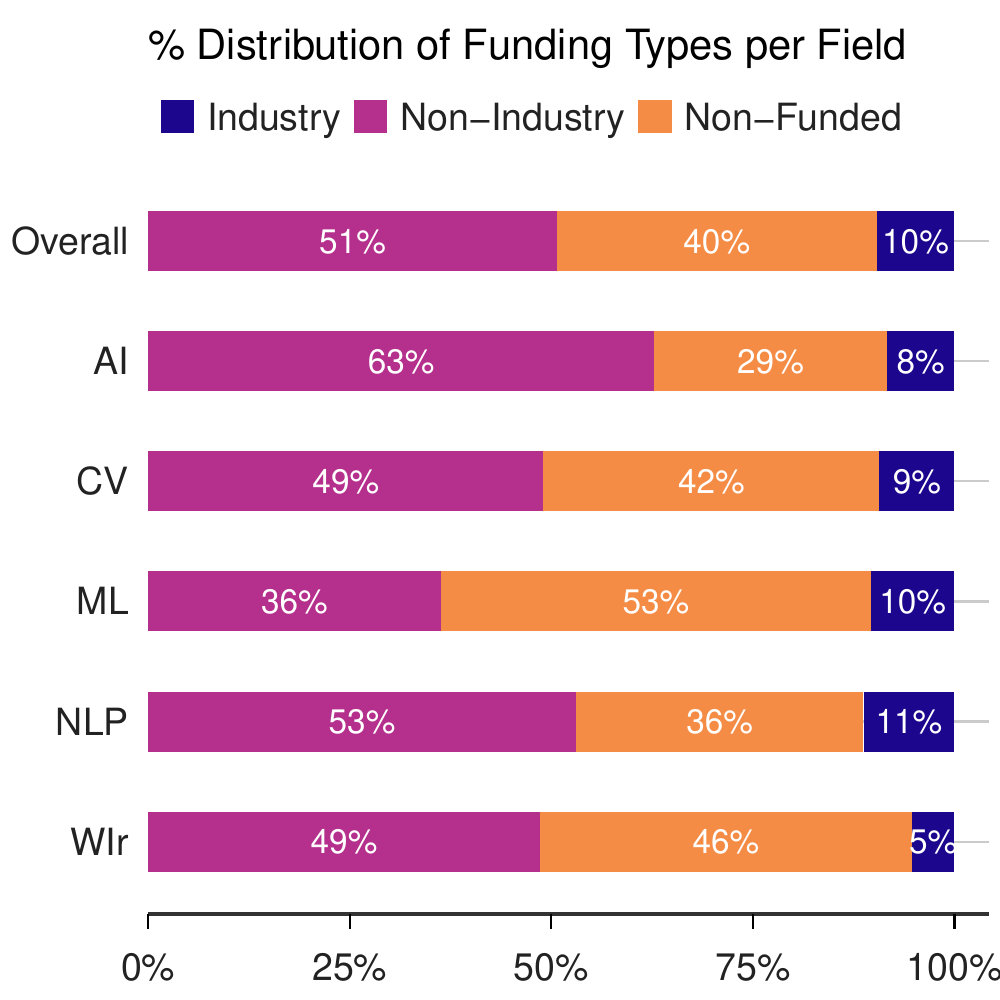}
        \caption{}
        \label{fig:fundingPercentageFundingTypePerField}
    \end{subfigure}
    \hfill
    \begin{subfigure}[ht]{.48\textwidth}
    \centering
        \includegraphics[scale = 0.36]{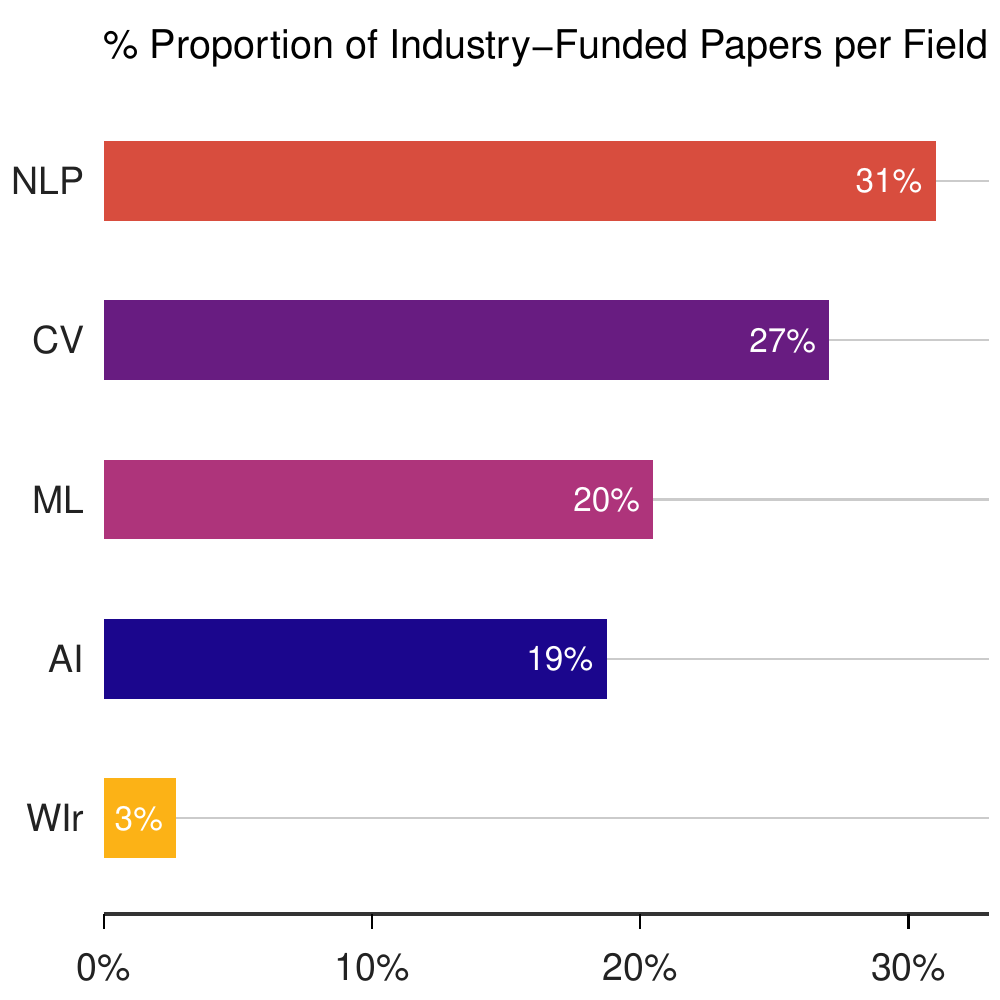}
        \caption{}
        \label{fig:fundingPercentagePerFieldTotal}
    \end{subfigure}
    \caption{The percentage distribution of (\subref{fig:fundingPercentageFundingTypePerField}) all funding types and (\subref{fig:fundingPercentagePerFieldTotal}) industry-funded papers for AI subfields.}
    \label{fig:fundingPercentagePerField}
    \vspace*{-.45cm}
\end{figure*}

\noindent \textbf{Results.}
\Cref{fig:fundingPercentageYears} tracks the evolution of industry-funded papers in AI research.
\Cref{fig:subFundingPercentageOverall_98_23} shows the long-term perspective on industry-funded AI research, while \Cref{fig:subFundingPercentageOverall_18_23} narrows the focus to the selected time frame of \num{2018}--\num{2022}. 
The proportion of industry-funded AI research increases markedly, from \SI{0.6}{\percent} in \num{1998} to \SI{9}{\percent} in \num{2022} (\Cref{fig:subFundingPercentageOverall_98_23}). 
A sharp growth occurs between 2016 (\SI{4}{\percent}) and \num{2020} (\SI{11}{\percent}), reflecting a \SI{180}{\percent} increase. 
However, our results show a decline post-\num{2020}, with the share of industry-funded papers dropping from \SI{11}{\percent} in \num{2020} to \SI{9}{\percent} in \num{2022}.  
Between \num{2018} and \num{2022}, \SI{10}{\percent} of all papers published at top AI conferences were funded by Big Tech (\Cref{fig:fundingPercentageFundingTypePerField}). 
\Cref{fig:fundingPercentagePerFieldTotal} shows the percentage of industry-funded papers in each subfield out of all industry-funded papers. 
NLP (\SI{31}{\percent}) and CV (\SI{27}{\percent}) have the largest share of industry-funded publications in Scopus.
Observe that despite its growth, WIr (\SI{3}{\percent}) lags behind. 
Overall, \SI{61}{\percent} of papers are funded, with Big Tech providing funding for \SI{10}{\percent} of papers on average (\Cref{fig:fundingPercentageFundingTypePerField}).
Non-profit or public organizations averagely fund \SI{51}{\percent} of papers.
In ML, \SI{36}{\percent} of papers are funded by non-profit or public organizations, and \SI{53}{\percent} of are not funded, while all other fields (i.e., industry-funded and non-industry-funded) have a funding percentage above \SI{50}{\percent}. 
The highest concentration of Big Tech-funded research occurs in NLP (\SI{11}{\percent}) and ML (\SI{10}{\percent}).

\noindent \textbf{Discussion.}
The results align with \cite{ahmed2023growing}’s observations of increasing industry presence in AI between \num{2016} and \num{2020}. However, our analysis reveals lower percentage values due to our high-precision method of focusing on acknowledging direct funding rather than author institution affiliations. 
Funding shapes research directions directly \cite{thelwall2023research}, indicating industry influence more precisely than affiliations alone.

The decline in industry funding post-\num{2020} does not necessarily imply a reduced industry presence in AI research. 
Preprint platforms such as arXiv can be confounding factors and show a non-linear increase in AI-related papers \cite{Krenn_2023,maslej2023artificialintelligenceindexreport}. 
Yet our observations suggest that the industry is shifting away from traditional conference publications in favour of less costly and time-consuming alternatives.
For example, the technical report for LLaMA \cite{touvron2023llamaopenefficientfoundation} and ChatGPT\footnote{\url{https://chat.openai.com}} was released only on arXiv and corporate website, respectively. 
The industry's ability to conduct exclusive research --- owing to their access to critical resources --- explains why these papers remain highly relevant despite avoiding peer-reviewed venues. 
The COVID-19 pandemic further accelerated the use of preprint servers as a rapid publication medium \cite{smart2022evolutionPrePrints}, while macroeconomic pressures such as supply chain disruptions, plummeting revenues, and rampant inflation have driven companies to cut costs \cite{jPMorganFutureBigTech2023}, possibly influencing their reduced investment in academic conference publications. 
The potential trend of industry migration from publishing at top AI conferences brings several challenges. 
This trend may lead to less correction of errors through errata and retractions \cite{smart2022evolutionPrePrints}.
It also may result in diminished conference sponsorship, as Big Tech may now be independent of conference publications to gain attention.

The high proportion of funded papers shown in \Cref{fig:fundingPercentageFundingTypePerField} --- particularly in AI (\SI{73}{\percent}) and NLP (\SI{64}{\percent}) --- highlights the financial demands of cutting-edge AI research, exceeding what unfunded individuals or groups can typically pay for. 
Public and non-profit institutions are the primary supporters of AI research, although the rising share of industry funding in specific fields reflects a growing corporate interest in the field's commercial applications. 
However, since Scopus bases funding information on paper acknowledgments, the true industry presence is likely underreported, as disclosures are sometimes omitted or forgotten \cite{wang2015there}. 
Furthermore, funding for AI research extends beyond the flow of money. 
Funding can include other benefits such as access to models, datasets, computational power, and expertise \cite{verdegem2024dismantling}.

\noindent \textbf{Q2.} \label{sec:Q2}
\textit{Papers from which funding sources --- industry (Big Tech), public, or non-profit ---a re most prominently cited in AI research papers? How has this citation trend evolved over time?}

\noindent \textbf{Ans.}
As we know from Q1, \SI{9}{\percent} of all papers published between \num{2018} and \num{2022} were funded by Big Tech. 
Another key question is whether industry-funded research attracts more citations than non-industry-funded and non-funded papers. 
Industry ownership of key resources may increase the visibility of these studies, even if their quantity is low. 
For a fair comparison, we consider the volume of funded papers by funding types (i.e., \textit{industry}-funded, \textit{non-industry}-funded, and \textit{non-funded}) and normalize the citation data according to each funding type's size. 
We introduce a new metric, called the \textit{Citation Preference Ratio} ($CPR$), which measures whether a paper with a specific funding type is cited more or less frequently than expected, based on its availability. 
A higher CPR indicates that a paper with a specific funding type is cited more often than the volume of papers would suggest.
The CPR from AI to a funding type $f$ is defined as follows:

\begin{align}
CPR_{AI}(f) &= \frac{C(f)}{E(f)} \label{eq:CPR} \\
\text{where } C(f) &= \sum_{\forall f_i \in F} C^{f_i \rightarrow f}, \\
\text{and } E(f) &= \left( \sum_{\forall f_i \in F} \sum_{\forall f_j \in F} C^{f_i \rightarrow f_j} \right) \cdot \frac{N_f}{N}
\end{align}

where $C(f)$ is the number of citations from all funding types to funding type $f$, $E(f)$ is the expected number of citations to $f$ proportional to its share of total papers, $C^{f_i \rightarrow f_j}$ is the number of citations from funding type $f_i$ to funding type $f_j$, $F$ is the set of all subfields, and $N$ is number of papers across all funding types. 
A $CPR>1$ shows that the funding type $f$ is more often cited by AI than expected; $CPR<1$ shows that AI less often cites the funding type $f$ than expected; and $CPR=1$ shows that the citations are proportional to availability (no citation preference).

\begin{figure*}[tbp]
    \centering
  \begin{subfigure}[ht]{.48\textwidth}
  \centering
  \includegraphics[scale = 0.38]{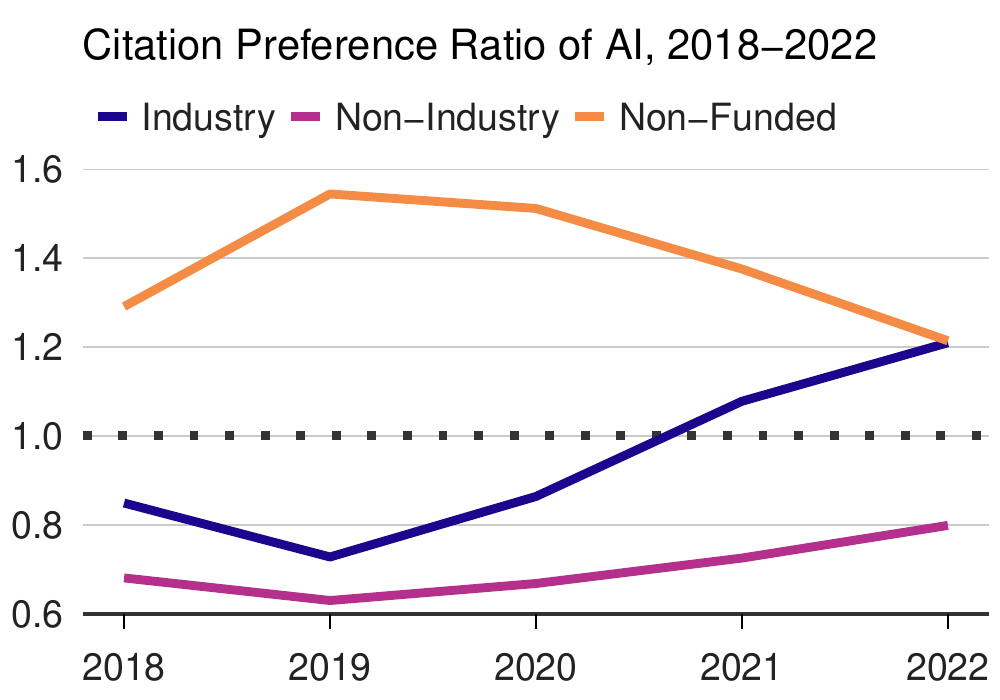}
  \caption{}
  
  \label{fig:CBR}
\end{subfigure} %
\hfill
\begin{subfigure}[ht]{.48\textwidth}
  \centering
  \includegraphics[scale = 0.5]{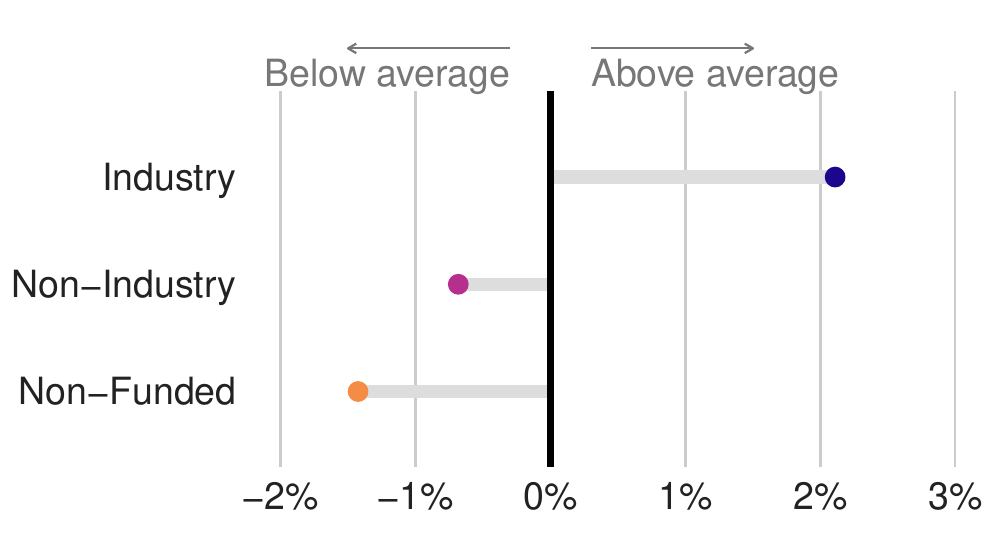}
  \caption{}

  \label{fig:ORCP}

\end{subfigure} %
\caption{(\subref{fig:CBR}) The \textit{CPR} of AI. (\subref{fig:ORCP}) Industry-funded research's \textit{ORCP} towards industry-funded, non-industry-funded, and non-funded papers. In both cases, higher values indicate more insularity.}
\vspace*{-.45cm}
\end{figure*}

\noindent \textbf{Results.}
\Cref{fig:CBR} reveals a consistent upward trend in citation preference towards funded papers since \num{2019}. 
By \num{2021}, the AI community started citing more industry-funded papers than expected by the number of papers.
Despite this growing trend, non-funded research was cited more frequently than industry-funded until \num{2022}. 
However, non-funded research's CPR has declined since \num{2019}, demonstrating that the AI community is citing fewer non-funded papers than expected by the growth.
Notably, the CPR of non-industry-funded papers remains low. 
Although there are growing numbers of papers annually, top AI papers cite non-funded papers disproportionately less. 
In contrast, they cite industry-funded papers disproportionately more than the growth in papers would suggest.

\noindent \textbf{Discussion.}
The increasing CPR for industry-funded and non-industry-funded papers demonstrates a link between funding and academic visibility in AI research. 
Voices in the community suggest that industry-funded research of Big Tech labs often serves as a label for excellent methodology and implementation.
The marked increase in citation preference of industry-funded research can also be attributed to the industry's contribution of tools and resources, such as frameworks, datasets, and models, which become foundational in AI research and development \cite{ahmed2020dedemocratization, ahmed2023growing, verdegem2024dismantling}. 
The publications of these tools are frequently cited in academic and industry research alike. 
An example is the paper introducing PyTorch, a widely used machine learning library created by Meta \cite{paszke2019pytorchimperativestylehighperformance}. %

In response to the growing relevance of AI, government agencies and non-profit organizations have increased their funding \cite{andre2024ai}. 
For example, the top \num{3} funding agencies in our citation graph based on occurrences were the EU Horizon 2020 Program (\SI{3}{\percent} of occurrences), the Engineering and Physical Sciences Research Council (\SI{1.8}{\percent} of occurrences), and the German Research Foundation (\SI{1.8}{\percent} of occurrences). 
This shift may have increased the quantity and quality of non-industry-funded publications, leading to rising citation counts. 
Our results show that research without external funding lags behind in citation metrics. 

\noindent \textbf{Q3.} \label{sec:Q3} 
\textit{To what extent do industry-funded papers cite other industry-funded papers as opposed to other funding types?}

\noindent \textbf{Ans.}
The growing presence of industry funding in AI research raises questions about potential self-reinforcing cycles, where industry-funded research may disproportionately cite other industry-funded work, potentially creating echo chambers~\cite{cinelli2021echo} that could shape research narratives and discourse even without explicit intention.
We calculate the difference in industry-funded outgoing citation percentage to a funding type $f$ versus the average outgoing citations from various funding types to $f$. 
We rely on the \textit{Outgoing Relative Citational Prominence} ($ORCP$) metric by \cite{wahle2023we} with one key modification: we adjust the notion of a paper being in specific research fields to a paper being funded by specific funding types (details in \Cref{ap:orcp}). 
If industry-funded research ($IF$) has an ORCP greater than \num{0} for $f$, then $IF$ cites $f$ more often than other funding types cite $f$ on average.

\noindent \textbf{Results.}
\Cref{fig:ORCP} shows the ORCP scores of industry-funded papers across funding types, with industry-funded research citing other industry-funded research more than average ($ORCP=2\%$).
Notably, despite the presence of extensive non-industry-funded and non-funded research of comparable quantity, both of these funding types have an $ORCP<0$, implying that industry-funded research cites non-industry-funded and non-funded work markedly less than how much the other funding types cite non-industry-funded and non-funded research.
Among all funding types, the highest ORCP to a funding type occurs within the same funding type, indicating that citations to papers of the same funding type are more common than cross-type citations. 
Non-industry-funded and non-funded research has higher ORCP scores to itself than industry-funded work has to itself. 
Additionally, industry-funded research has the lowest ORCP among the three funding types when cited by non-industry-funded and non-funded papers.

\noindent \textbf{Discussion.}
The findings show that papers from one funding type prefer to cite papers from the same funding type than others.
Despite this pattern, the degree of insularity remains moderate, reflected by an ORCP of \SI{2}{\percent}. 
Citation insularity emerges from a complex interplay of factors beyond mere research specialization. 
While industry and public funding sources encompass diverse organizations with varied research priorities, several structural factors may contribute to self-referential citation practices. 
These can include shared methodological frameworks and established collaborative networks for industry-funded research.
Similarly, despite their diverse missions, public funding agencies often create interconnected research communities through targeted programs and review processes, potentially reinforcing specific citation patterns.

\textbf{Q4.} \label{sec:Q4}
\textit{How well are industry-funded papers cited? How does the citation impact vary between different funding types?}

\noindent \textbf{Ans.}
We examine the citation impact of industry-funded papers as a measure of influence on other researchers. 
We analyze median citations, mean citations, and the h5-index \cite{hirsch2005index} across different funding types. 
The h5-index is a proxy for impact and influence despite its known limitations in capturing all research dimensions \cite{bornmann2007we, costas2007h}.

\noindent \textbf{Results.}
\Cref{tab:citationImpact} shows the mean and median number of citations and the h5-index for all papers of a funding type.
Observe how funded papers have the highest mean citations (industry-funded: \num{304.07}; non-industry-funded: \num{170.68}), median citations (industry-funded: \num{86}; non-industry-funded: \num{44}), and h5-index (industry-funded: \num{570}; non-industry-funded: \num{893}). 
Non-industry-funded research has the highest h5-index (\num{893}) and the largest volume of papers (\num{25190}), demonstrating a substantial number of highly cited papers and total contributions. Conversely, industry-funded research achieves a disproportionately high h5-index (\num{570}) relative to the number of papers, reflecting a focus on high-yield research outputs.
By comparison, unfunded research ranks lowest in citation impact among all funding types. 
Industry-funded research stands out, with \SI{12}{\percent} of papers having high citation impact, compared to only \SI{4}{\percent} of non-industry-funded and \SI{2}{\percent} of non-funded papers.
However, industry-funded show consistent citation patterns, suggesting a steady impact compared to other funding types, which tend to have more one-hit successes.

We conducted Levene's test to assess the relationship between funding type and citation counts.
The results indicate a weak positive correlation ($\rho = 0.133$), which is highly statistically significant ($p < 0.0001$). 
While this finding confirms a correlation between funding type and citation counts, it also highlights that funding type accounts for only \SI{13.3}{\percent} of the observed variance in citation counts. 
This observation emphasizes the importance of additional factors, such as research quality, topical relevance, and author reputation.%

\begin{table}[tbp]
\small
    \caption{The total number of AI papers published in the last five years, the median and mean number of citations, and the h5-index for different funding types (by decreasing h5-index).}
    \label{tab:citationImpact}
        \begin{tabular}{lcccc}
            \toprule
            \textbf{Funding Type} & \textbf{Count} & \textbf{Median} & \textbf{Mean} & \textbf{h5-index} ($\downarrow$) \\
            \midrule
            Non-Ind.-Funded & \num{25190} & \num{44} & \num{170.68} & \num{893} \\ 
            Ind.-funded & \num{4801} & \num{86} & \num{304.07} & \num{570} \\ 
            Non-Funded & \num{19820} & \num{10} & \num{48.86} & \num{380} \\  
            \bottomrule
        \end{tabular}
    \vspace*{-.55cm}
\end{table}

\noindent \textbf{Discussion.}
\sloppy
The number of citations to funded papers shows a connection between research funding and citations. %
Big Tech-funded research has a disproportionately high citation impact relative to its publication volume compared to other types, a phenomenon attributable to multiple underlying mechanisms. 
A possible reason is that industry often recruits students and faculties, which brings established citation networks, increasing visibility and engagement.
However, a high citation count is not necessarily an indicator of research excellence; citations are only one factor. 
A balanced public and private funding approach is essential for fostering a robust and equitable AI research ecosystem. 
Metrics beyond citations (e.g., transparency) help evaluate research impact, emphasize reproducibility, societal relevance, and ethical considerations.
Public institutions can encourage industry participation in open science while providing researchers with the resources necessary to make decisions about engaging in or opting out of industry collaborations.

\noindent \textbf{Q5.}
\textit{Which fields do industry-funded papers cite? How diverse are the outgoing citations in these papers?}

\noindent \textbf{Ans.}
We analyze outgoing citations by funding type to determine whether non-industry-funded and non-funded research concentrates on industry-favored topics. 
We calculate each funding type's share of citations directed to various fields, defined as the percentage of citations to a given field from a given funding type over all citations from a given funding type to any field. 
For papers associated with multiple fields, each field receives a citation. 
We use the fields Scopus pre-classification system\footnote{This classification is done by Scopus internally by experts using the All Science Journal Classification scheme \cite{scopus2024ASJC}}, which categorizes a paper based on the aims and scope of the title and its content. 

\begin{figure}
    \centering
    \includegraphics[width=\columnwidth]{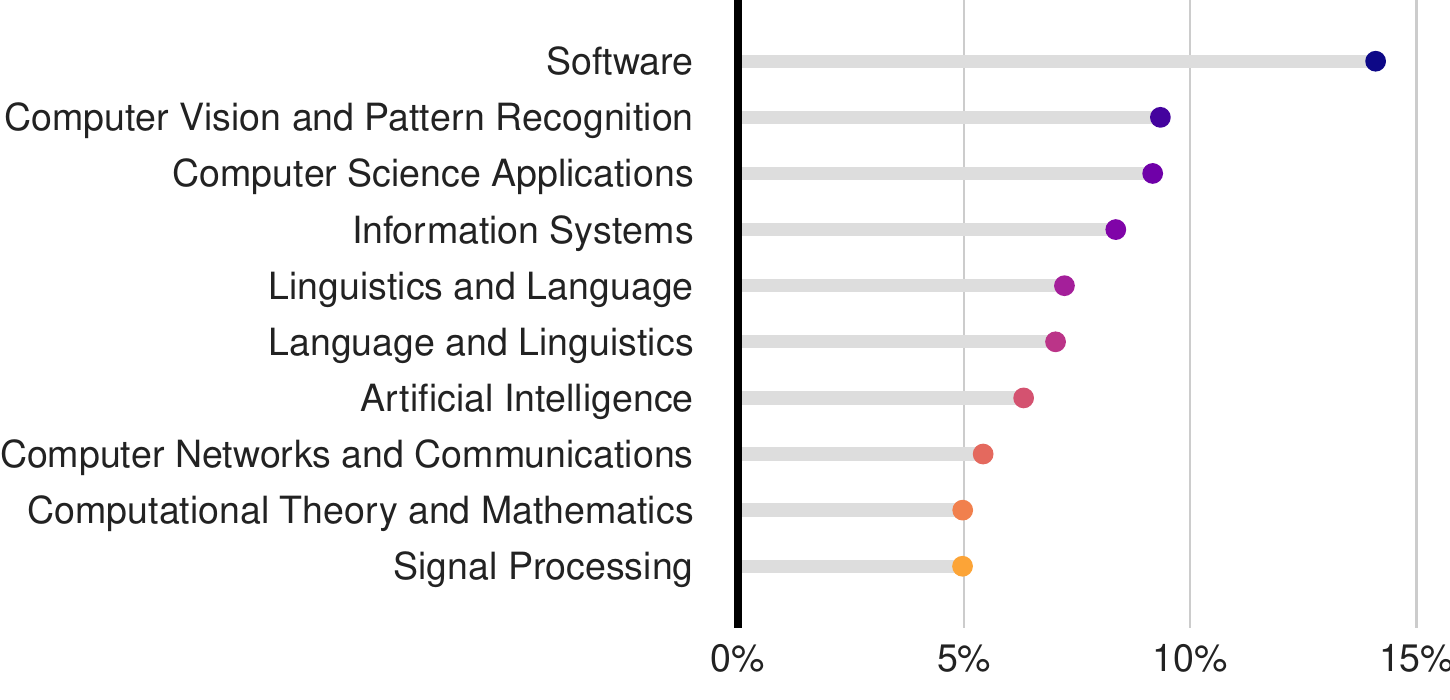}
    \caption{Percentage of outgoing citations from industry-funded papers to the top \num{10} most cited fields.}
    \label{fig:citingFieldIndustry}
    \vspace*{-.45cm}
\end{figure}

\noindent \textbf{Results.}
\Cref{fig:citingFieldIndustry} shows the distribution of citations to the top \num{10} cited fields for industry-funded papers. %
Across all funding types, \num{8} out of \num{10} most cited fields belong to computer science, highlighting a strong focus on this field and a low outgoing citation field diversity. 
The top \num{10} fields account for over \SI{70}{\percent} of outgoing citations in each funding type. %
Industry-funded research shows the highest concentration, with \SI{77}{\percent} of citations directed to these top \num{10} fields, compared to \SI{75}{\percent} and \SI{72}{\percent} for non-industry-funded and non-funded papers, respectively.
The \num{4} main fields cited remain consistent across funding types, constituting more than \SI{50}{\percent} of citations within the top \num{10} fields, indicating a common primary interest across funding types. 
However, industry-funded papers show an increased interest in linguistics, while non-industry-funded and non-funded papers emphasize AI more prominently. 
Additionally, non-industry-funded research shows a stronger orientation toward theoretical work, contrasting with the industry and non-funded paper's emphasis on networks and signal processing.

\noindent \textbf{Discussion.}
The convergence of research fields across funding types, alongside the growing engagement with industry-funded work, demonstrates Big Tech influence. %
Building on the findings of \cite{klinger2022narrowingairesearch}, we show that industry-funded research exhibits lower thematic diversity compared to non-industry-funded and non-funded research, demonstrated by the high citation density in the \num{10} ten most-cited fields. 
Our results reflect a concentration of industry-funded work in data-hungry and computationally intensive fields, such as computer vision and information retrieval (information systems). 
In contrast, research not funded by Big Tech shows a relatively strong focus on AI and theoretical fields such as mathematics. 
This emphasis does not preclude engagement with data-driven or computationally intensive fields. 
Instead, it indicates a broader distribution of research interests compared to the narrower focus of industry-funded work.

\noindent \textbf{Q6.}
\textit{What is the average age of cited papers for different funding types?}

\noindent \textbf{Ans.}
Scholars frequently engage with related work across disciplines to validate or challenge earlier findings, situate their contributions, and extend the boundaries of knowledge. 
However, the tendency to not cite enough relevant good work from the past (more than a few years old) --- referred to as ``citation amnesia'' \cite{singh2023forgotten} --- remains a pressing issue. 
Forgetting some older works can be helpful, as it makes space for new ideas.
However, too much forgetting can lead to unnecessary reinvention of methods.
Building on insights from \cite{singh2023forgotten,wahle2024citation}, we investigate temporal citation patterns to quantify this phenomenon.
We analyze the citations for each funding type to other papers and calculate how far back in time the \textit{cited} papers are published. 
When a paper $x$ cites a paper $y_i$, then the age of the citation ($AoC$) is the difference between the year of publication of $x$ and $y_i$, as in $AoC(x, y_i) = YoP(x) - YoP(y_i)$.
We then calculate the $AoC$ for each citation in a paper and average them (\textit{mAoC}).

\begin{figure}[tbp]
  \centering
  \includegraphics[width=\columnwidth]{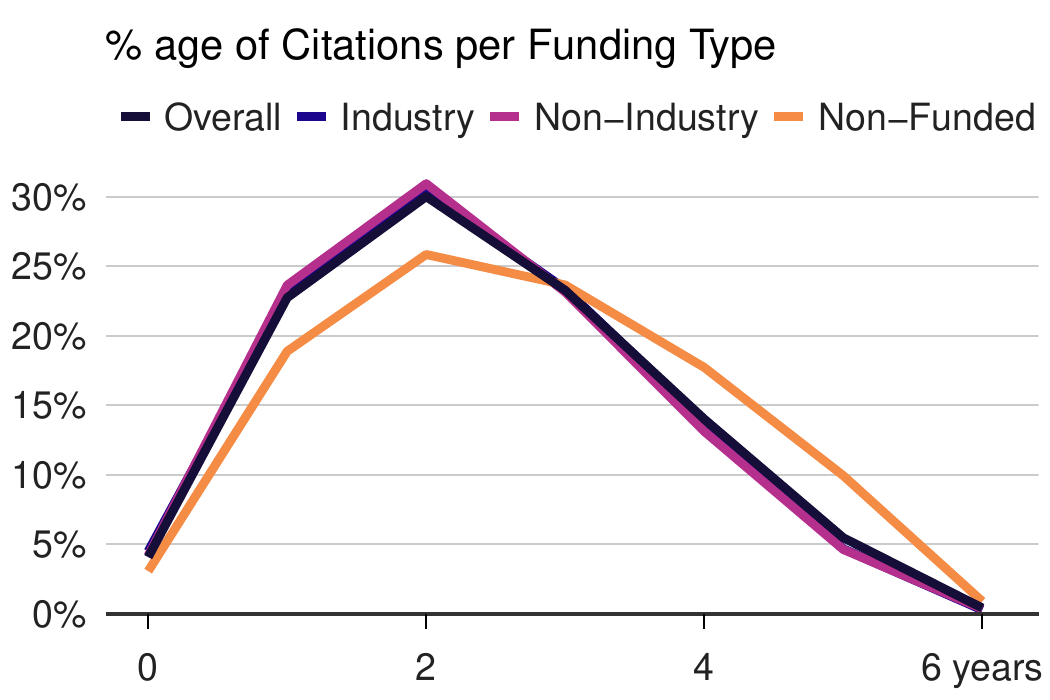}
  \caption{Distribution of citation age as measured by \textit{AoC} for papers in AI (overall and by funding type).}
  \vspace*{-.45cm}
  \label{fig:AoCFundingType}

\end{figure}

\noindent \textbf{Results.}
\Cref{fig:AoCFundingType} shows the distribution of \textit{mAoCs} for all papers of each funding type and overall across the years after the publication of the \textit{cited} paper. 
The y-axis represents the average percentage of citations papers received for the years (x-axis).
Most citations occur for papers published \num{2} years prior ($AoC=2$). 
The citation patterns for all funding types show a similar increasing trend, followed by an abrupt decline in the following years. 
Non-funded papers have a lower peak value but maintain higher citation rates than other funding types. 
Additionally, industry-funded papers have the highest percentage of citations in the publication year, while non-funded papers have the lowest.
The \textit{mAoC} for papers published between \num{2018} and \num{2022} shows industry-funded papers have the lowest mean \textit{mAoC} of \num{4.79}, followed closely by non-industry-funded papers with \num{4.92}, and non-funded papers at \num{5.03} (details in \Cref{tab:mAoc} in \Cref{ap:mAoc}).

\textbf{Discussion.}
Our results show that papers typically receive the highest number of citations \num{2} years after publication, followed by a strong decline, which is consistent with related work \cite{singh2023forgotten, wahle2024citation}. 
At their peak, non-funded papers receive fewer citations than funded research, but their citation decline is more gradual. 
This dynamic, coupled with the higher \textit{mAoC} value (\Cref{tab:mAoc}), suggests that non-funded research accumulates citations over a longer time and cites older work compared to funded research.
Non-funded work may focus on citing foundational or theoretical contributions that continue attracting citations over time. 
In contrast, funded research may prioritize emerging topics and cutting-edge innovations.
This encourages new developments to build upon recent advances, often neglecting foundational work.%

\noindent \textbf{Q7.}
\textit{What is the distribution of \textit{mAoC} in industry-funded papers? How does this distribution vary over the years?}

\noindent \textbf{Ans.}
We compute the \textit{mAoC} for each industry-funded paper and the median and mean \textit{mAoC} for all such papers over time.
\Cref{ap:mAoc} provides more details on the \textit{mAoC}.

\begin{figure}[tbp]
  \centering
  \includegraphics[width=\columnwidth]{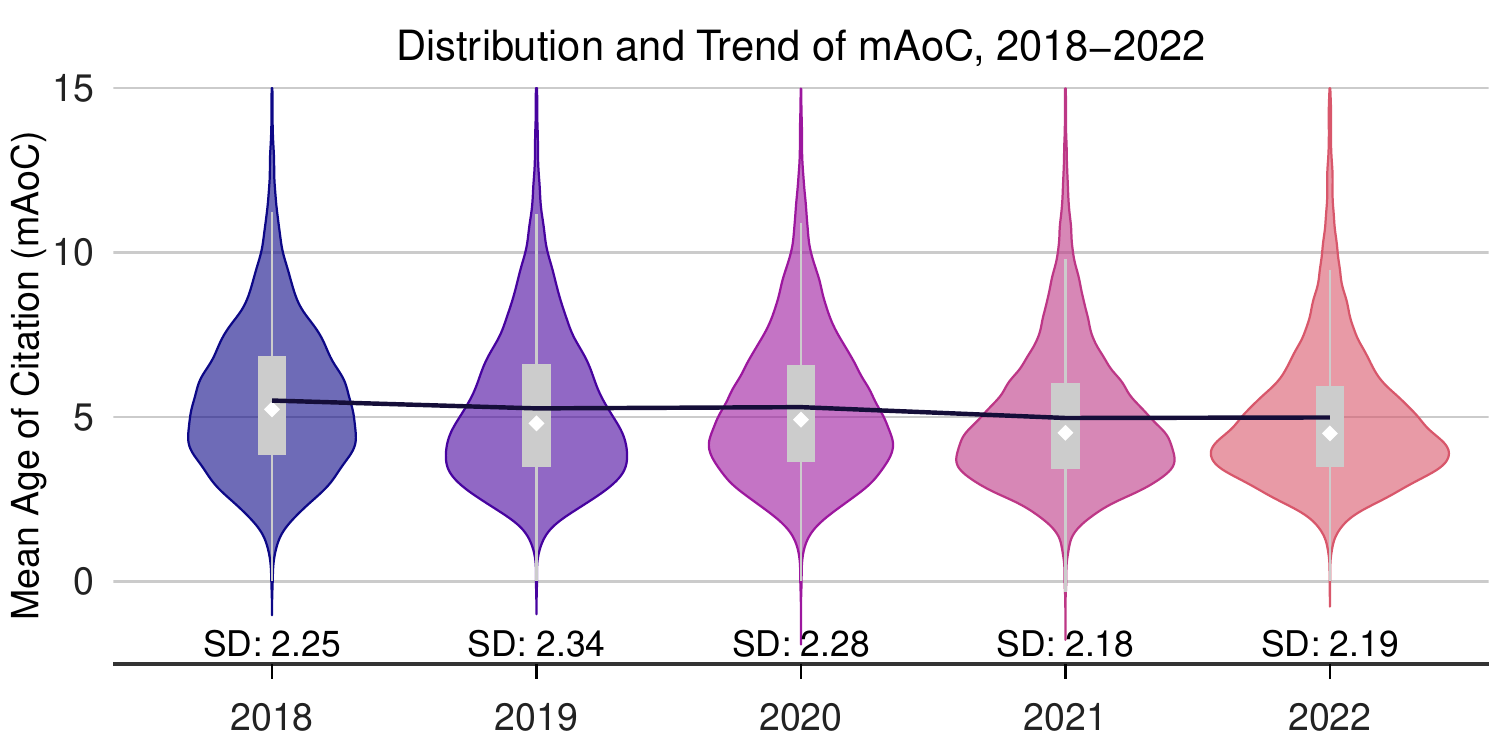}
  \caption{Citation ages as measured by \textit{mAoC} for industry-funded papers between \num{2018} and \num{2022}. The standard deviation for each year is displayed below the respective violin plot. The median (white diamond) and the mean \textit{mAoC} (dark line) are shown for each year.}
  \vspace*{-.4cm}
  \label{fig:mAoCOverTime}

\end{figure}

\noindent \textbf{Results.}
\Cref{fig:mAoCOverTime} shows the violin plots for distributions of \textit{mAoC} in industry-funded papers across years. 
Each plot highlights the median (white diamond within the grey rectangle), representing the recency of citations for that year. 
Over time, the median \textit{mAoC} for industry-funded papers shows a consistent decline, indicating an increasing focus on recent work. 
The shrinking size of the second and third quartiles (halves of the grey rectangle) indicates that citations are increasingly concentrated around the median, reflecting a narrowed citation age.
From \num{2018} to \num{2022}, the mean \textit{mAoC} closely follows the median trend, but it remains consistently higher, revealing a right skew in the data. 
This skew is due to several papers citing much older papers. %
The decreasing standard deviation also suggests diminishing citation age diversity, possibly reinforcing the trend toward citing newer literature.
By \num{2022}, the violin's density transforms into a spinning tractroid top\footnote{Form of the iconic spinning top in the movie Inception.}, reflecting a concentration of \textit{mAoC} values near five years.%

\noindent \textbf{Discussion.}
The results show a decline in citations to older works within industry-funded papers and a reduction in the temporal diversity of citations. 
Although the exact causes of this trend remain uncertain, multiple factors contribute to the evolving citation dynamics. 
The substantial impact of transformers on NLP and ML can increasingly favour more recent publications. 
The pressures of the ``publish or perish'' paradigm further exacerbate this trend, encouraging researchers to divide their work into smaller, publishable units \cite{nature_publish_2010}. 
Our results add to and are consistent with the mean-citation age results in \cite{wahle2024citation}.%
Our analysis focuses on AI publications between \num{2018} and \num{2022}, situating these results in the overall trajectory of how temporal citation patterns in AI have evolved recently.%

\section{Concluding Remarks}

This work examined the impact of Big Tech funding on AI research. 
To enable this analysis, we compiled a dataset that includes $\approx$\num{49.8}\,K AI papers, their funding agencies, citations from other papers to AI papers, and citations from AI papers to other papers. 
We introduced the \textit{Citation Preference Ratio}, a novel metric that shows a growing trend within the AI research community to reference Big Tech-funded work.
We also employed established metrics such as the \textit{Relative Citational Prominence}, which highlights an increasing insularity in industry-funded research, and the \textit{Mean Age of Citation} metric, which shows the tendency of industry-funded research to cite recent literature.
While the presence of Big Tech-funded research in top AI conferences is declining, its citational influence continues to grow. 
Industry-funded papers cite fewer contributions from non-industry-funded and non-funded research despite a more extensive growth in paper volume than industry-funded papers (growing insularity). 

The manual and automated methods for identifying funding agencies in our work introduce risks of interpretation and matching errors.
Relying solely on citation-based metrics captures influence but cannot fully explain the qualitative drivers behind these trends.
A broader analysis of industry impact, encompassing resource allocation and the interplay between public and private funding, are still necessary.
From an ethical perspective, this work underscores the importance of interpreting these citation-based findings responsibly, remembering that metrics alone cannot dictate value, nor should they justify prioritizing one funding source over another.

As researchers, we are not only observers of these trends and fast paced developments. 
Instead, we have agency in this process. 
One could further say we, as a field, have a responsibility to reflect on these trends, to discuss and vote for appropriate actions to shape and direct the field's future. 
Public institutions play a critical role in this effort by improving their funding policies and providing researchers with the tools and support they need to receive diverse project funding.
Bridges between industry and public funding --- to learn from one another and benefit from infrastructure --- are vital for a healthy research environment.
Government efforts to provide competitive infrastructure, such as computing, data, and compensation, are also key to attracting talent for open research.

\bibliographystyle{IEEEtran}
\bibliography{source}

\newpage
\appendix

\section{Appendix} \label{sec:Appendix}

\subsection{Details of the Limitations} \label{sec:A_limitations}

\subsubsection{Manual Analysis}

The manual analysis has a few limitations. 
First, because examining thousands of funding agencies individually is time-intensive, this analysis only includes \SI{5}{\percent} of the extracted funding agencies. 
Notably, this \SI{5}{\percent} covers \SI{74}{\percent} of all funding occurrences, providing robust overall representation. 
Second, we identified IFs based on available online information. 
In cases where insufficient data prevented us from confirming a IF, we marked agencies as non-funded, potentially leading to false negatives in the dataset's metadata. 
Third, since this analysis was performed solely by one person, interpretation variability may affect the consistency and quality of the data.

\subsubsection{Automatic Analysis}

Automated fuzzy text matching with standardized company names can incorrectly match unrelated agencies, leading to false positives when identifying IFs. 
To mitigate this risk of false positives, a high similarity threshold (\SI{90}{\percent}) was set, although this conservative threshold could reduce the number of matches and potentially miss some IFs. 
However, most funding agency names included the standardized identifier (e.g., Google, Google DeepMind, Google Cloud), minimizing the risk of missing relevant IFs.

\subsubsection{General Limitations}

Beyond technical limitations, this study faces broader constraints. 
Industry involvement in AI research today extends beyond direct financial contributions to include access to models, datasets, computational resources, and specialized expertise \cite{montes2019distributed,riedl2020ai,ahmed2023growing, verdegem2024dismantling}. 
This analysis relies exclusively on Scopus funding data, which originates from paper acknowledgments~\cite{liu2020accuracy} to capture Big Tech influence. 
The extent of funding information in these acknowledgments varies depending on the details provided in the publications, reflecting disciplinary and regional disparities in funding reporting practices \cite{pranckutėScopus21}. 
Notably, studies by \cite{liu2020accuracy} and \cite{pranckutėScopus21} identify errors in Scopus funding acknowledgment text and funding agency fields. 
It is crucial to interpret the consistency and quality of the identified industry presence with a grain of salt.

This research focuses on publications from prominent AI-related conferences rather than all AI-related academic publications. 
Although leading conferences shape the academic research agenda \cite{freyna2010journalstatus}, this selection excludes vibrant, often non-English AI communities and venues, limiting the generalizability of our findings to the global AI research landscape. 
Future studies must explore industry influence across diverse sub-communities and venues worldwide.

Furthermore, this analysis quantifies influence primarily through citations, a method with inherent limitations. 
Citation counts lack nuance, as not all citations reflect the same level of influence \cite{valenzuela2015identifying,zhu2015measuring}. 
Additionally, citation patterns are affected by biases \cite{mohammad-2020-examining,ioannidis2019standardized,nielsen2021global}. 
This work also examines citation practices on a large scale, focusing on quantitative trends. Qualitative aspects may reveal why Big Tech-funded research receives more engagement within the AI community, shows growing insularity, and cites recent over older literature. 
Several factors may contribute to this, such as the volume of recent publications, the applicability of industry-funded research, and the technical relevance of industry-funded work.

Our analysis also did not address the allocation of financial resources by industry across AI subfields and conferences. 
Analyzing the cash flows from industry to AI research and their impact over time could reveal whether financial resources markedly drive influential AI research or whether other resources are key. 
We leave the exploration of cash flows, their impact, and their presence over time for future work.

\subsection{Details of Ethical Considerations} \label{sec:A_EthicalConsiderations}

This study analyzes the scientific literature at an aggregate level and not on individual papers or authors, using data from the Scopus database. 
The database provides metadata such as titles, authors, funding agencies, and publication years, which are used without infringing copyrighted content. 

A critical aspect of this study is its reliance on citation counts as a proxy to characterize funding types. 
Although citations serve as a convenient metric, this approach raises concerns about potential misinterpretation or misuse of our findings. 
For example, the observed high number of outgoing citations to Big Tech-funded research should not be used as a rationale for diminishing research funded by non-industry sources or conducted without external funding. 
A more comprehensive evaluation framework may be beneficial in addressing the risks of oversimplified interpretations. 
Such a framework would integrate multiple dimensions, including relevance, popularity, resource availability, impact, geographic context, and temporal trends, thus mitigating the problems of shallow analysis.

\subsection{Details on the Extraction of Companies} \label{sec:A_DEC}

We searched for company names and common aliases (e.g., Microsoft, Microsoft Azure, Microsoft Cloud Computing Research Centre) using the fuzzywuzzy python package\footnote{\url{https://pypi.org/project/fuzzywuzzy/}} with a \SI{90}{\percent} threshold. 

\begin{table}[h]
\small
    \caption{Company name standardization.}
    \label{tab:CompanyStandard}
        \begin{tabularx}{\linewidth}{Xlr}
            \toprule
            \textbf{Names of funding agencies}  &  &   \textbf{Std. Name} \\
            \midrule
            Microsoft, Microsoft Azure, Microsoft Research &  & Microsoft \\ 
            Amazon, AWS, Amazon Research &  & Amazon \\
            Google, Google DeepMind, Google Cloud &  & Google \\
            Nvidia, NVIDIA AI Center, NVIDIA Corp &  & Nvidia \\
        \bottomrule
        \end{tabularx}
    \caption{The table shows examples of standardized funding agency names.}
\end{table}

\subsection{Details on the Outgoing Relative Citational Prominence (ORCP)} \label{ap:orcp}

We rely on the \textit{Outgoing Relative Citational Prominence} ($ORCP$) metric by \cite{wahle2023we} with one key modification: we adjust the notion of a paper being in specific research fields to a paper being funded by specific funding types. 
If industry-funded research ($IF$) has an ORCP greater than \num{0} for $f$, then $IF$ cites $f$ more often than other funding types cite $f$ on average.
The following equation shows that metric.

\begin{align}
ORCP_{IF}(f) &= X(f) - Y(f)
\label{eq:ORCP}\\
\text{where } X(f) &= \frac{C^{IF \rightarrow f}}{\sum_{\forall f_j \in F} C^{IF \rightarrow f_j}},\\
\text{and } Y(f) &= \frac{1}{N} \sum_{i=1}^N \frac{C^{f_i \rightarrow f}}{\sum_{\forall f_j \in F} C^{f_i \rightarrow f_j}} 
\end{align}

where $F$ is the set of all funding types, $N$ is the number of all funding types, i.e. \num{3}, and $C^{f_i \rightarrow f_j}$ represents the number of citations from papers in funding type $f_i$ to papers in funding type $f_j$.

\subsection{Details on the mean Age of Citations} \label{ap:mAoc}

The age of the citation ($AoC$) is the difference between the year of publication ($YoP$) of $x$ and $y_i$:

\begin{align}
AoC(x, y_i) = YoP(x) - YoP(y_i)
\label{eq:AoC}
\end{align}

We then calculate the AoC for each of the citations of a paper and average them:

\begin{align}
mAoC(x, y_i) = \frac{1}{N} \sum_i^N AoC(x, y_i)
\label{eq:mAoC}
\end{align}

where $N$ denotes the total number of references in paper $x$.
For example, if a paper $x$ from 2022 cites two papers, one from 2010 and one from 2020, the $mAoC$ of the paper $x$ is \num{7}~years.

\Cref{tab:mAoc} shows the mean \textit{mAoC} for papers published between \num{2018} and \num{2022}, grouped by funding type. 
Observe how industry-funded papers have the lowest mean \textit{mAoC} of \num{4.79}, followed closely by non-industry-funded papers with a mean \textit{mAoC} of \num{4.92}, and non-funded papers at \num{5.03}.

\begin{table}[htbp]
    \centering
    \label{tab:mAoc}
        \begin{tabular}{lr}
            \toprule
            \textbf{Funding Type} &  \textbf{\textit{mAoC} $\pm$ 95\% Conf.} ($\uparrow$)\\
            \midrule
            Industry &  \num{4.79} $\pm$ \num{0.02} \\
            Non-Industry &  \num{4.92} $\pm$ \num{0.01} \\  
            Non-Funded &  \num{5.08} $\pm$ \num{0.02} \\  
        \bottomrule
        \end{tabular}
    \caption{The \textit{mAoC} and confidence intervals for different funding types are ordered by increasing \textit{mAoC}.}
\end{table}

\subsection{Supplementary Citation Graph Details} \label{sec:A_ZG}

\begin{figure}[htbp]
  \centering
  \includegraphics[width = \columnwidth]{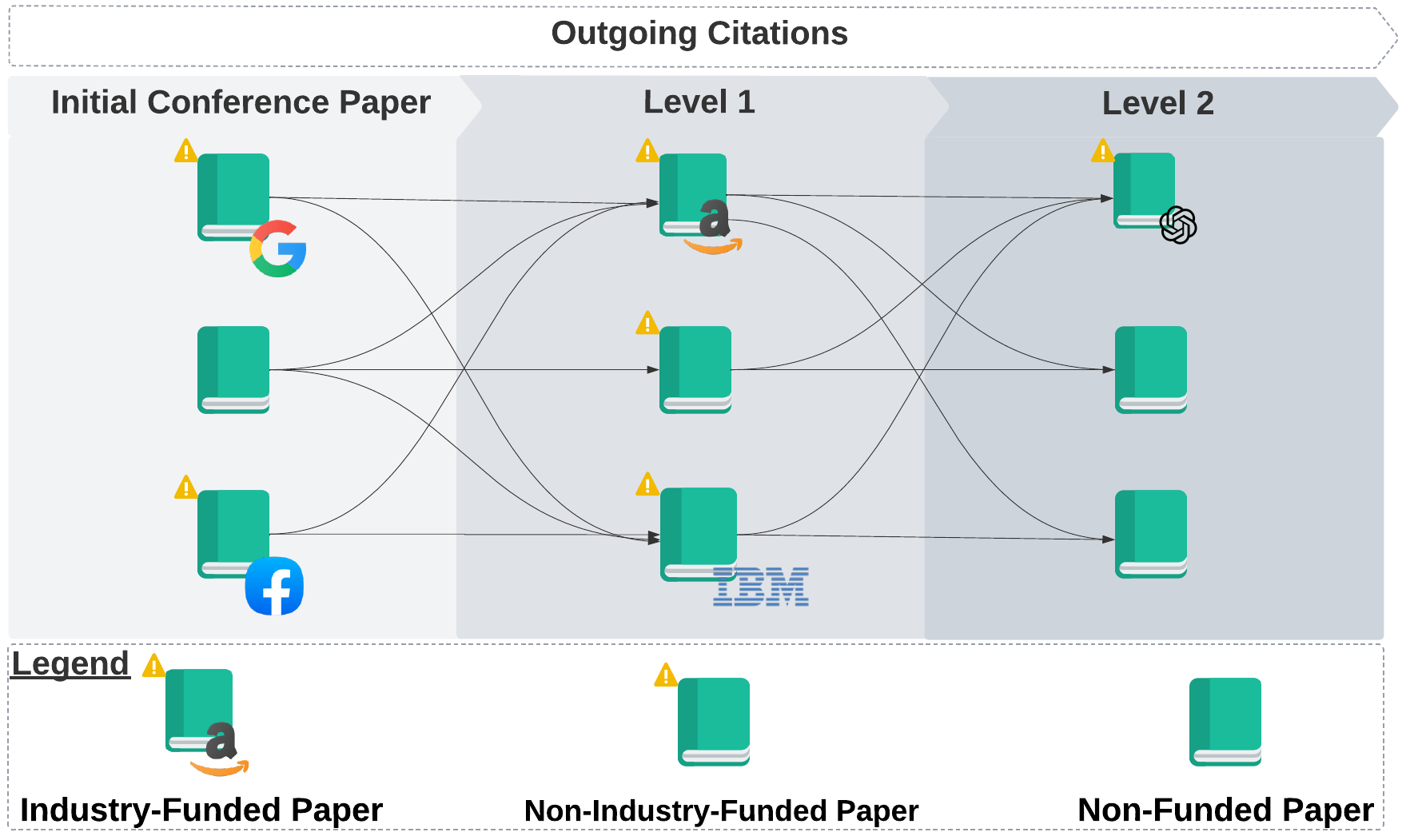}
  \caption{Example citation graph to identify funding agencies for the search scope.}
  \label{fig:citationGraph}
\end{figure}

\Cref{fig:citationGraph} shows an sample of the described citation graph in \Cref{sec:DC}.

\subsection{Full Conference Names} \label{sec:A_FCN}

\Cref{tab:fullConferenceNames} shows the full names of the matched and unmatched top AI conferences.

\begin{table}[htbp]
\small
    \label{tab:fullConferenceNames}
        \begin{tabularx}{\linewidth}{Xlcc}
            \toprule
            \textbf{Full Name} & & \textbf{Acronym} & \textbf{Field} \\
            \midrule
            Advancement of Artificial Intelligence & & AAAI & \multirow{2}{*}{AI} \\ 
            International Joint Conference on Artificial Intelligence & & IJCAI & \\
            \addlinespace
            Conference on Computer Vision and Pattern Recognition & & CVPR & \multirow{3}{*}{CV} \\ 
            International Conference on Computer Vision* & & ICCV* & \\
            European Conference on Computer Vision$^{\dag}$ & & $\text{ECCV}^{\dag}$ & \\
            \addlinespace
            International Conference on Machine Learning* & & ICML* & \multirow{3}{*}{ML} \\ 
            International Conference on Learning Representations & & ICLR & \\
            Conference and Workshop on Neural Information Processing Systems$^{\dag}$ & & $\text{NeurIPS}^{\dag}$ & \\
            \addlinespace
            Association for Computational Linguistics & & ACL & \multirow{2}{*}{NLP} \\ 
            Empirical Methods in Natural Language Processing & & EMNLP & \\
            \addlinespace
            International Conference on Web Search and Data Mining* & & WSDM* & \multirow{3}{*}{WIr} \\ 
            Conference on Research and Development in Information Retrieval & & SIGIR & \\
            $\text{International World Wide Web Conferences}^{\dag}$ & & $\text{WWW}^{\dag}$ & \\
            \bottomrule
            \addlinespace
            \multicolumn{3}{@{}l}{\footnotesize *Replacing conference. $^{\dag}\text{Replaced conference.}$}
        \end{tabularx}
        \caption{Full names, acronym, and field of matched and unmatched AI conferences.}
\end{table}

\subsection{Supplemental Experimental Results} \label{sec:A_SER}

In addition to the primary results presented in this study, we describe supplementary results in the form of additional statistics and plots.

\subsubsection{Extended Results on Citation Preference Ratio}\label{sec:A_ERCPR}
\begin{figure*}[htbp]
    \begin{subfigure}[ht]{.3\textwidth}
        \centering
        \includegraphics[scale = 0.3]{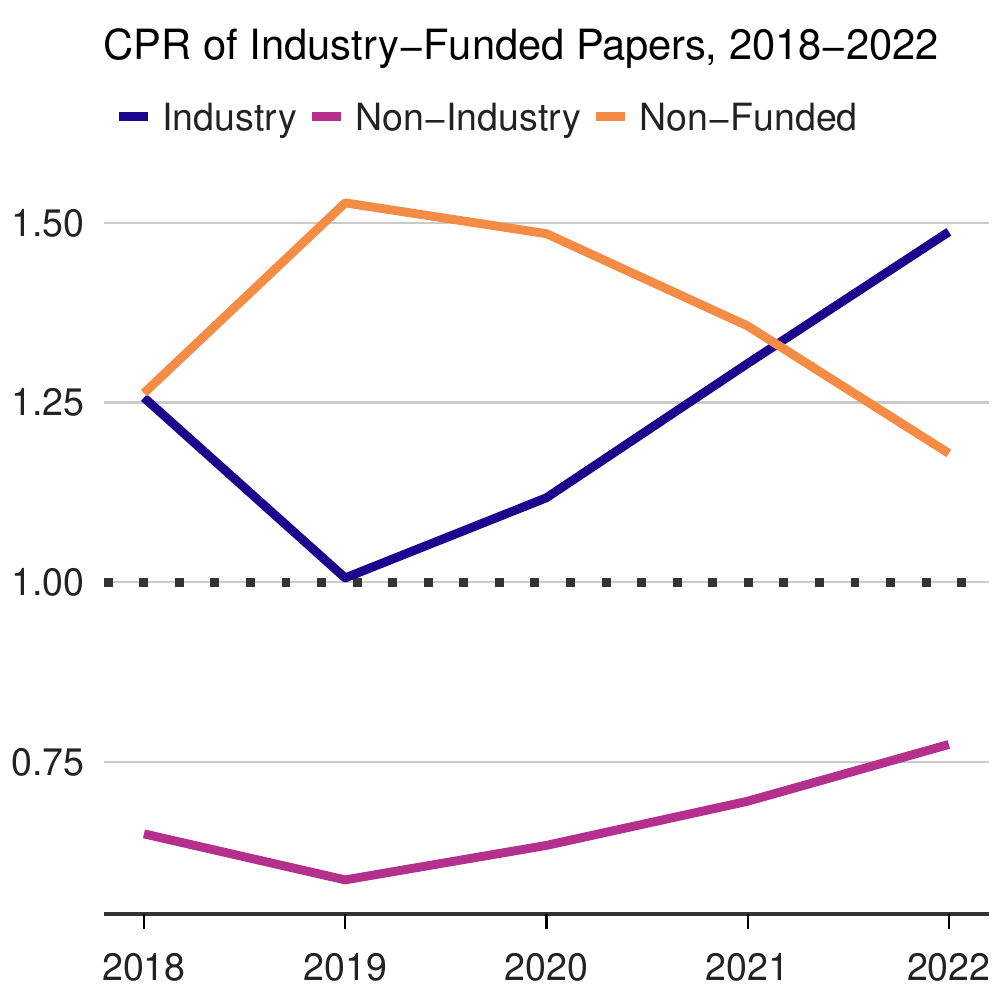}
        \caption{}
        \label{fig:subCPRIndustry}
    \end{subfigure}
    \hfill
    \begin{subfigure}[ht]{.3\textwidth}
        \centering
        \includegraphics[scale = 0.3]{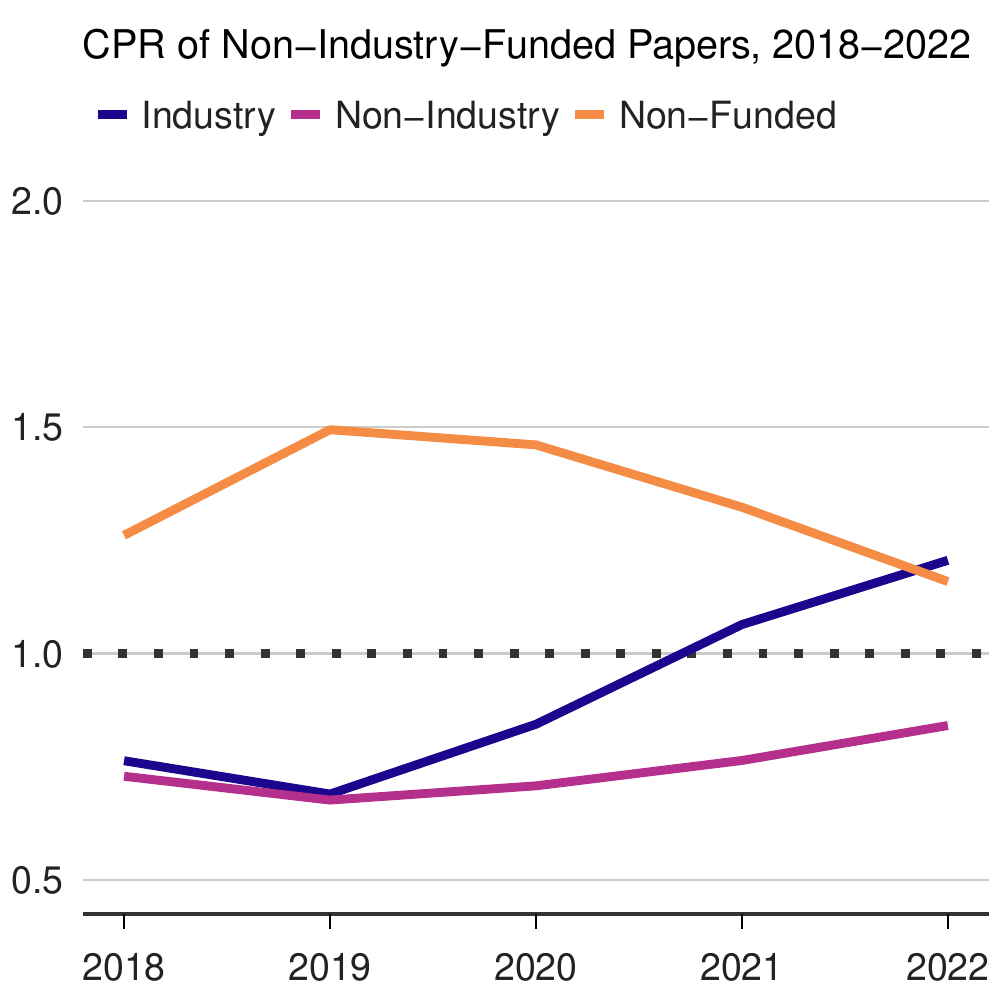}
        \caption{}
        \label{fig:subCPRNonIndustry}
    \end{subfigure}
    \hfill
    \begin{subfigure}[ht]{.3\textwidth}
        \centering
        \includegraphics[scale = 0.3]{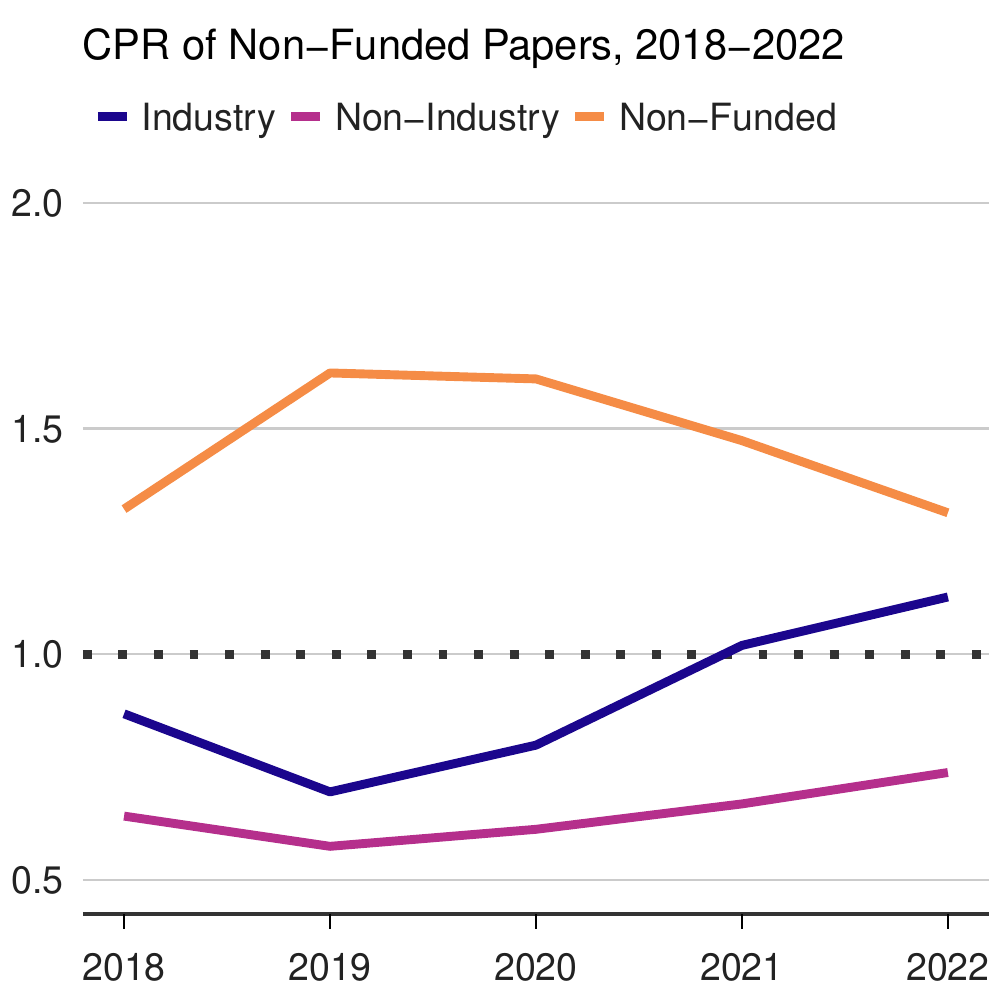}
        \caption{}
        \label{fig:subCPRNonFunded}
    \end{subfigure}
    \caption{The Citation Preference Ratio (CPR) of (\subref{fig:subCPRIndustry}) industry-funded papers, (\subref{fig:subCPRNonIndustry}) non-industry-funded papers, and (\subref{fig:subCPRNonFunded}) non-funded papers towards all funding types over time.}
    \label{fig:A_CPR}
\end{figure*}

\leavevmode\newline
\Cref{fig:A_CPR} shows the Citation Preference Ratio (CPR) of industry-funded papers, non-industry-funded papers, and non-funded papers to different funding types over time.

\subsubsection{Extended Results on Outgoing Relative Citation Prominence}\label{sec:A_ERRCP}
\begin{figure*}[htbp]
    \begin{subfigure}[ht]{.45\textwidth}
        \centering
        \includegraphics[scale = 0.4]{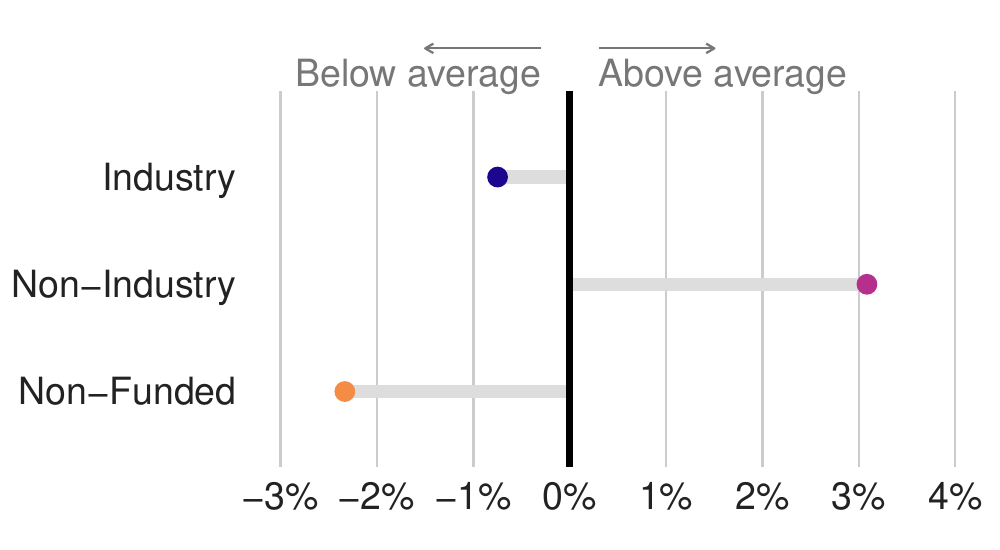}
        \caption{}
        \label{fig:subORCPNonIndustry}
    \end{subfigure}
    \hfill
    \begin{subfigure}[ht]{.45\textwidth}
        \centering
        \includegraphics[scale = 0.4]{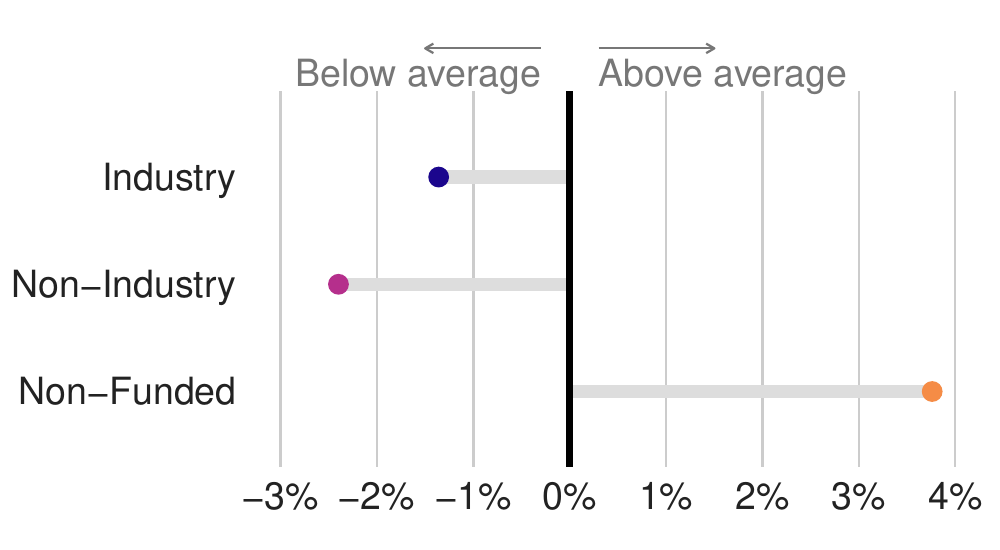}
        \caption{}
        \label{fig:subORCPNonFunded}
    \end{subfigure}
    \caption{Outgoing Relative Citational Prominence (ORCP) scores for (\subref{fig:subORCPNonIndustry}) non-industry-funded papers and (\subref{fig:subORCPNonFunded}) non-funded papers.}
    \label{fig:A_ORCP}
\end{figure*}

\leavevmode\newline
\Cref{fig:A_ORCP} shows the ORCP for (\subref{fig:subORCPNonIndustry}) non-industry-funded papers, and (\subref{fig:subORCPNonFunded}) non-funded papers.

\subsubsection{Extended Results on Cited Fields}\label{sec:A_ERCF}
\begin{figure*}[htbp]
    \begin{subfigure}[t]{.45\textwidth}
        \centering
        \includegraphics[scale = 0.31]{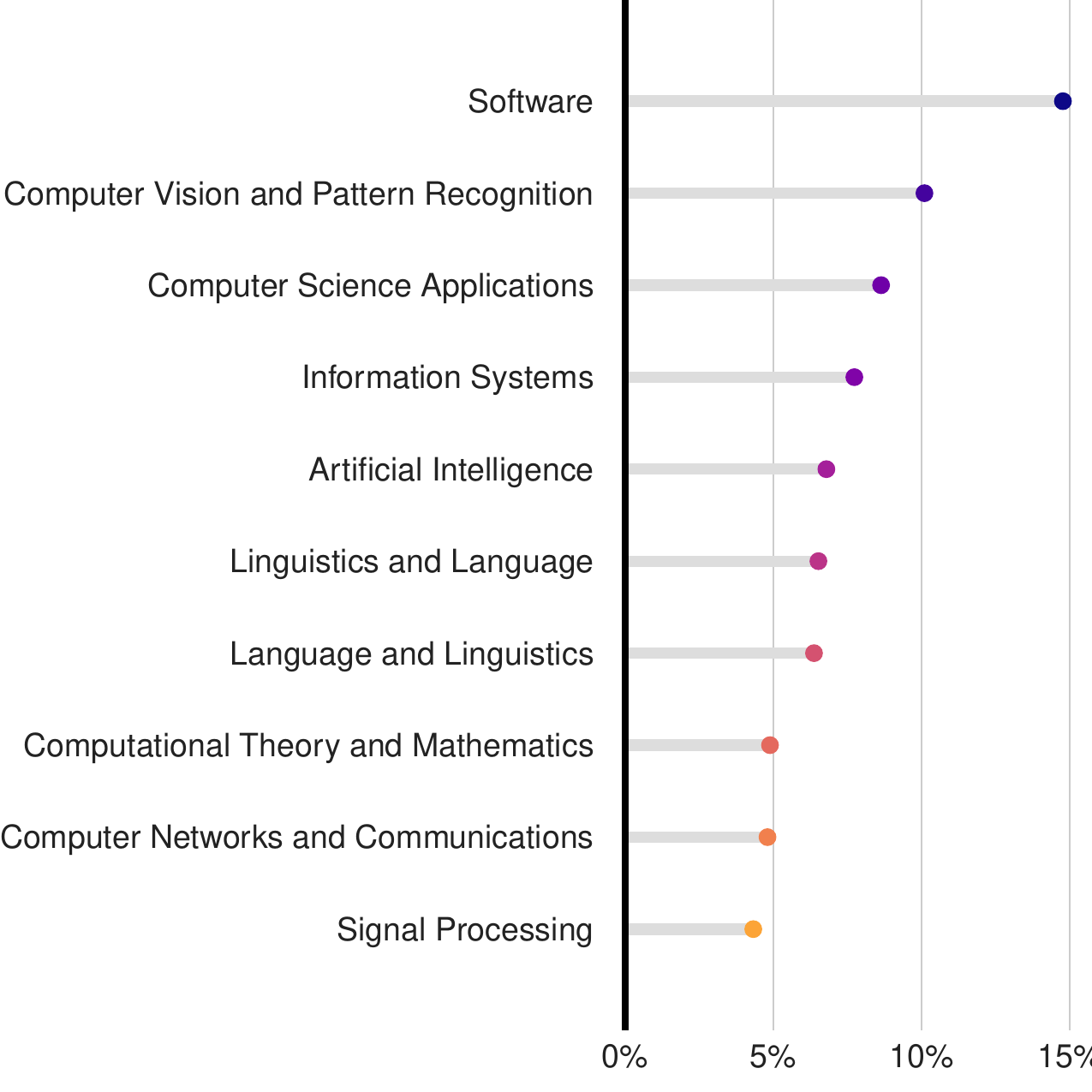}
        \caption{}
        \label{fig:subCitingFieldNonIndustry}
    \end{subfigure}
    \hfill
    \begin{subfigure}[t]{.45\textwidth}
        \centering
        \includegraphics[scale = 0.31]{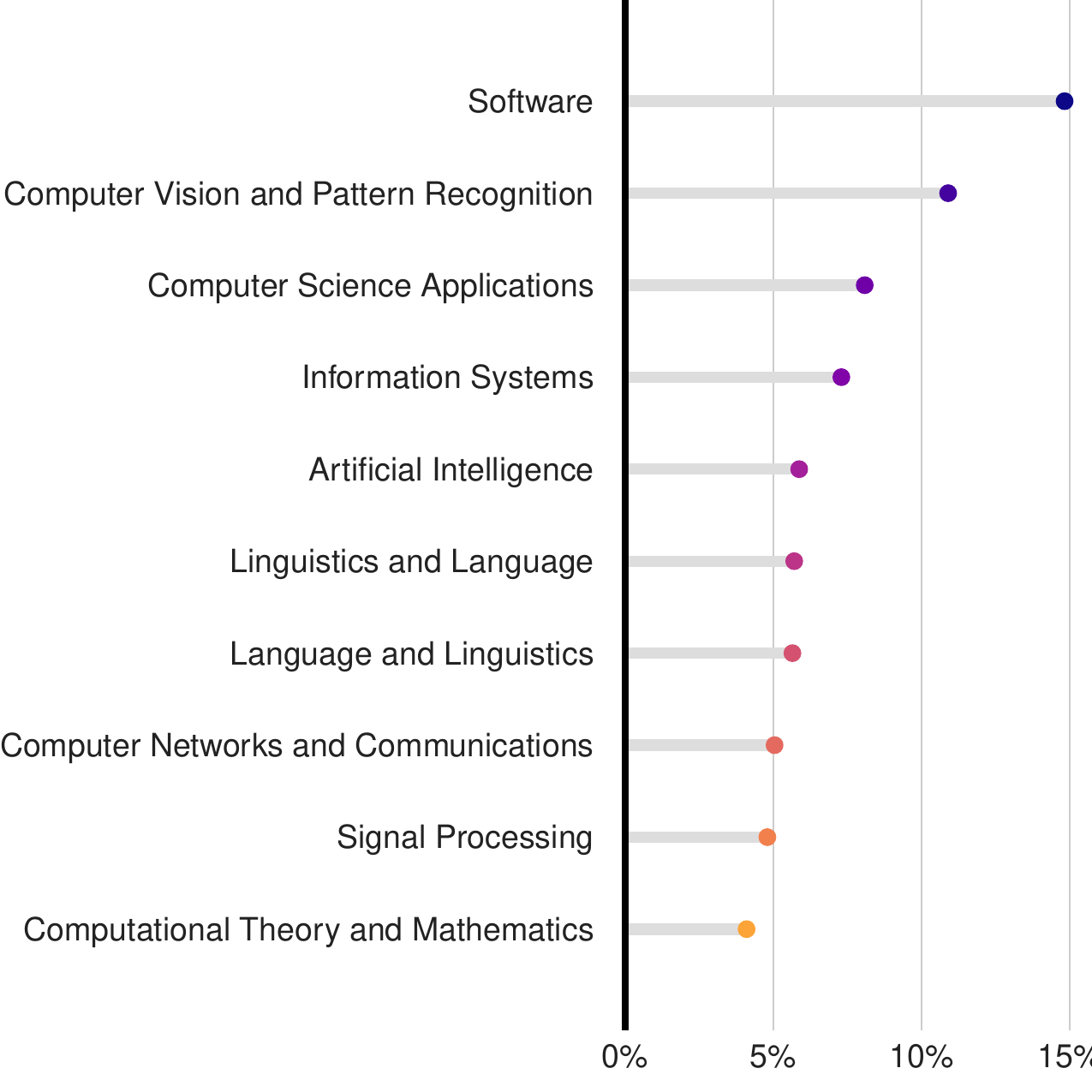}
        \caption{}
        \label{fig:subCitingFieldNonFunded}
    \end{subfigure}
    \caption{Percentage of outgoing citations from non-industry-funded papers (\subref{fig:subCitingFieldNonIndustry}) and non-funded papers (\subref{fig:subCitingFieldNonFunded}) to the top \num{10} cited fields.}
    \label{fig:citingFields}
\end{figure*}

\leavevmode\newline
\Cref{fig:citingFields} shows the top \num{10} cited fields for (\subref{fig:subCitingFieldNonIndustry}) non-industry-funded papers and (\subref{fig:subCitingFieldNonFunded}) non-funded papers.

\subsection{Additional Research Questions}\label{app:srq1}

\noindent \textbf{ARQ1.} \label{sec:Q5}
\textit{Which type of funding is most influenced by industry-funded research? How has this changed over the years?}

\noindent \textbf{Ans.}
To determine the funding types most affected by industry-funded research, we analyze the citation sources to industry-funded research by funding type. 
Thus, we calculate the average percentage of industry-funded references per paper, i.e., the mean ratio of citations from papers with a given funding type to industry-funded papers relative to the total citations of papers with that funding type. 
This approach measures how much different funding types rely on or interact with industry-funded research over time.

\begin{figure}[htbp]
  \centering
  \includegraphics[scale = 0.48]{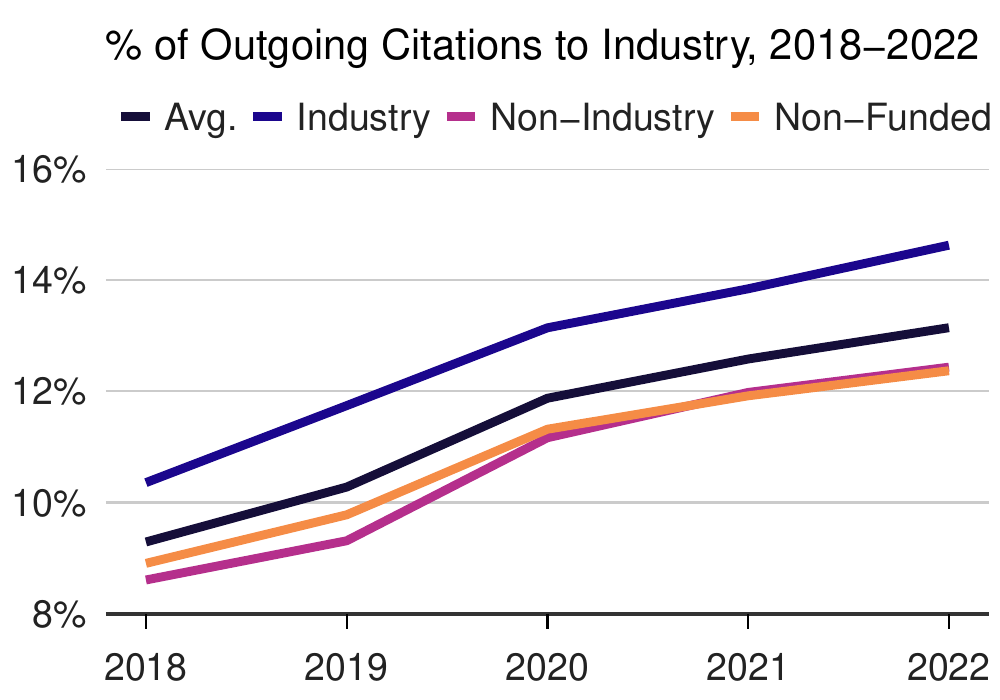}
  \caption{The average proportion of industry-funded references across various funding types for each paper. The macro-average shows the mean percentage of industry-funded references per paper overall funding types.}
  \label{fig:OCIndustry}
\end{figure}

\noindent \textbf{Results.}
\Cref{fig:OCIndustry} shows the proportion of outgoing citations to industry-funded papers per paper and funding type over time. 
It also shows the macro average of this proportion across all funding types. 
The share of citations referencing Big Tech-funded research has increased markedly across all funding types since \num{2018}. 
Non-industry-funded papers showed strong growth, with \SI{44}{\percent} increase in citations to industry-funded work per paper between \num{2018} and \num{2022} after a lower percentage start. 
This growth surpasses the growth of industry-funded papers (\SI{41}{\percent}) and non-funded papers (\SI{39}{\percent}). 
Despite this rise in cross-funding-type engagement, industry-funded papers maintained the highest proportion of outgoing citations per paper to other industry-funded research (\SI{15}{\percent} in \num{2022}), underlining the self-referential trend of industry-funded research.

\noindent \textbf{Discussion.}
The rising proportion of outgoing citations to industry-funded papers highlights a marked increase in engagement with industry-funded research across funding types. 
Non-industry-funded researchers show growing interest in industry-driven topics and methodologies. 
It is still unclear why non-industry-funded research experienced a substantial increase in engagement with industry-funded work. 
One possible reason for this engagement is the collaboration between industry and non-industry entities, with industry often providing cutting-edge resources and academia providing the marketplace to identify and recruit talented researchers. 
\cite{klinger2022narrowingairesearch} supports this perspective by highlighting significant industry-academia collaborations in AI research. They caution that such partnerships may narrow thematic diversity in favour of industry-preferred topics.

\subsection{AI Usage Card}

We report how we used AI assistants such as ChatGPT and Claude for this work in the following standardized card according to \cite{wahle2023aiusagecardsresponsibly}.

\clearpage
\onecolumn

{\sffamily
    \centering
    \tcbset{colback=white!10!white}
    \begin{tcolorbox}[
        title={\large \textbf{AI Usage Card} \hfill \makebox{\qrcode[height=1cm]{https://ai-cards.org}}},
        breakable,
        boxrule=0.7pt,
        width=.8\paperwidth,
        center,
        skin=bicolor,
        before lower={\footnotesize{AI Usage Card v1.0 \hfill \url{https://ai-cards.org} \hfill \href{https://jpwahle.com/ai-cards-preprint}{PDF} | \href{https://jpwahle.com/cite/jcdl2023wahle.bib}{BibTeX}}},
        segmentation empty,
        halign lower=center,
        collower=white,
        colbacklower=tcbcolframe]
            
        \footnotesize{
            \begin{longtable}{p{.15\paperwidth} p{.275\paperwidth} p{.275\paperwidth}}
              {\color{LightBlue} \MakeUppercase{Correspondence(s)}} \newline Jan Philip Wahle
              & {\color{LightBlue} \MakeUppercase{Contact(s)}} \newline \href{mailto:wahle@uni-goettingen.de}{wahle@uni-goettingen.de}
              & {\color{LightBlue} \MakeUppercase{Affiliation(s)}} \newline University of Göttingen
              \\\\
              & {\color{LightBlue} \MakeUppercase{Project Name}} \newline Big Tech-Funded AI Papers Have Higher Citation Impact, Greater Insularity, and Larger Recency Bias 
              & {\color{LightBlue} \MakeUppercase{Key Application(s)}} \newline Scientometrics, Artificial Intelligence, Funding influence, Big Tech impact, Power asymmetries, Monopolization, Echo Chamber, Ethical AI
              \\\\
              {\color{LightBlue} \MakeUppercase{Model(s)}} \newline ChatGPT \newline Claude
              & {\color{LightBlue} \MakeUppercase{Date(s) Used}} \newline 2024-04-01 \newline 2024-05-01
              & {\color{LightBlue} \MakeUppercase{Version(s)}} \newline 4o, 4o1 \newline 3.5 Sonnet\\\\
              \cmidrule{2-3}\\
      
              {\color{LightBlue} \MakeUppercase{Ideation}} \newline    
              & {\color{LightBlue} \MakeUppercase{Generating ideas, outlines, and workflows}} \newline Not used 
              & {\color{LightBlue} \MakeUppercase{Improving existing ideas}} \newline Not used \\\\
              & {\color{LightBlue} \MakeUppercase{Finding gaps or compare aspects of ideas}} \newline Not used \\\\
              
              {\color{LightBlue} \MakeUppercase{Literature Review}} \newline    
              & {\color{LightBlue} \MakeUppercase{Finding literature}} \newline Not used
              & {\color{LightBlue} \MakeUppercase{Finding examples from known literature}} \newline Not used \\\\
              & {\color{LightBlue} \MakeUppercase{Adding additional literature for existing statements and facts}} \newline Not used
              & {\color{LightBlue} \MakeUppercase{Comparing literature}} \newline Not used \\\\
              \cmidrule{2-3}\\
      
              {\color{LightBlue} \MakeUppercase{Methodology}} \newline    
              & {\color{LightBlue} \MakeUppercase{Proposing new solutions to problems}} \newline Not used
              & {\color{LightBlue} \MakeUppercase{Finding iterative optimizations}} \newline Not used \\\\
              & {\color{LightBlue} \MakeUppercase{Comparing related solutions}} \newline Not used \\\\
              
              {\color{LightBlue} \MakeUppercase{Experiments}} \newline    
              & {\color{LightBlue} \MakeUppercase{Designing new experiments}} \newline Not used
              & {\color{LightBlue} \MakeUppercase{Editing existing experiments}} \newline Not used \\\\
              & {\color{LightBlue} \MakeUppercase{Finding, comparing, and aggregating results}} \newline Not used \\\\
              \cmidrule{2-3}\\
      
              {\color{LightBlue} \MakeUppercase{Writing}} \newline ChatGPT Claude  
              & {\color{LightBlue} \MakeUppercase{Generating new text based on instructions}} \newline Used
              & {\color{LightBlue} \MakeUppercase{Assisting in improving own content}} \newline Used \\\\
              & {\color{LightBlue} \MakeUppercase{Paraphrasing related work}} \newline Used 
              & {\color{LightBlue} \MakeUppercase{Putting other works in perspective}} \newline Not used \\\\
              
              {\color{LightBlue} \MakeUppercase{Presentation}} \newline    
              & {\color{LightBlue} \MakeUppercase{Generating new artifacts}} \newline Not used
              & {\color{LightBlue} \MakeUppercase{Improving the aesthetics of artifacts}} \newline Not used \\\\
              & {\color{LightBlue} \MakeUppercase{Finding relations between own or related artifacts}} \newline Not used \\\\
              \cmidrule{2-3}\\
              {\color{LightBlue} \MakeUppercase{Coding}} \newline ChatGPT Claude  
              & {\color{LightBlue} \MakeUppercase{Generating new code based on descriptions or existing code}} \newline Used
              & {\color{LightBlue} \MakeUppercase{Refactoring and optimizing existing code}} \newline Used \\\\
              & {\color{LightBlue} \MakeUppercase{Comparing aspects of existing code}} \newline Not used \\\\
              
              {\color{LightBlue} \MakeUppercase{Data}} \newline    
              & {\color{LightBlue} \MakeUppercase{Suggesting new sources for data collection}} \newline Not used 
              & {\color{LightBlue} \MakeUppercase{Cleaning, normalizing, or standardizing data}} \newline Not used  \\\\
              & {\color{LightBlue} \MakeUppercase{Finding relations between data and collection methods}} \newline Not used  \\\\
              \cmidrule{2-3}\\
      
              {\color{LightBlue} \MakeUppercase{Ethics}} \newline    
              & {\color{LightBlue} \MakeUppercase{What are the implications of using AI for this project?}} \newline Generating code and improving the clarity of writing the paper has improved the efficacy of performing this scientific work.
              & {\color{LightBlue} \MakeUppercase{What steps are we taking to mitigate errors of AI for this project?}} \newline We manually fact-checked generated texts and inspected source code for potential generated bugs. \\\\
              & {\color{LightBlue} \MakeUppercase{What steps are we taking to minimize the chance of harm or inappropriate use of AI for this project?}} \newline We did not include text suggestions that had any chance of impacting marginalized groups.
              & {\color{LightBlue} \MakeUppercase{The corresponding authors verify and agree with the modifications or generations of their  used AI-generated content}} \newline Yes \\
      
            \end{longtable}
        }
        \tcblower
    \end{tcolorbox}
}

\twocolumn

\clearpage
\onecolumn
\hypertarget{annotation}{}
\pagestyle{empty}
\lstset{
  basicstyle=\footnotesize\ttfamily,
  breaklines=true,
  breakatwhitespace=false,
  columns=flexible,
  numbers=none
}

\definecolor{Primary}{RGB}{59, 130, 246}    %
\definecolor{PrimaryDark}{RGB}{30, 64, 175} %
\definecolor{LightBg}{RGB}{239, 246, 255}   %
\definecolor{TextDark}{RGB}{31, 41, 55}     %
\definecolor{TextMuted}{RGB}{107, 114, 128} %

\begin{tikzpicture}[remember picture, overlay]
  \fill[Primary] ([xshift=0cm,yshift=0cm]current page.north west) rectangle ([xshift=\paperwidth,yshift=-0.4cm]current page.north west);
\end{tikzpicture}

\vspace{0.8cm}
\begin{center}
  {\fontsize{22}{26}\selectfont\sffamily\bfseries \textcolor{PrimaryDark}{CiteAssist}}\\[0.2em]
  {\Large\sffamily\scshape \textcolor{TextMuted}{Citation Sheet}}\\[0.8em]
  {\small\sffamily Generated with \href{https://citeassist.uni-goettingen.de/}{\textcolor{Primary}{\texttt{citeassist.uni-goettingen.de}}}
  \CiteAssistCite{}
  }\end{center}

\begin{center}
\vspace{1em}
\begin{tikzpicture}
\draw[Primary, line width=0.6pt] (0,0) -- (\textwidth,0);
\end{tikzpicture}
\vspace{1.2em}
\end{center}

\begin{tcolorbox}[enhanced,
                 frame hidden,
                 boxrule=0pt,
                 borderline west={2pt}{0pt}{Primary},
                 colback=LightBg,
                 sharp corners,
                 breakable,
                 fonttitle=\sffamily\bfseries\large,
                 coltitle=Primary,
                 title=BibTeX Entry,
                 attach title to upper={\vspace{0.2em}\par},
                 left=12pt]
\lstset{
    inputencoding = utf8,  %
    extendedchars = true,  %
    literate      =        %
      {á}{{\'a}}1  {é}{{\'e}}1  {í}{{\'i}}1 {ó}{{\'o}}1  {ú}{{\'u}}1
      {Á}{{\'A}}1  {É}{{\'E}}1  {Í}{{\'I}}1 {Ó}{{\'O}}1  {Ú}{{\'U}}1
      {à}{{\`a}}1  {è}{{\`e}}1  {ì}{{\`i}}1 {ò}{{\`o}}1  {ù}{{\`u}}1
      {À}{{\`A}}1  {È}{{\`E}}1  {Ì}{{\`I}}1 {Ò}{{\`O}}1  {Ù}{{\`U}}1
      {ä}{{\"a}}1  {ë}{{\"e}}1  {ï}{{\"i}}1 {ö}{{\"o}}1  {ü}{{\"u}}1
      {Ä}{{\"A}}1  {Ë}{{\"E}}1  {Ï}{{\"I}}1 {Ö}{{\"O}}1  {Ü}{{\"U}}1
      {â}{{\^a}}1  {ê}{{\^e}}1  {î}{{\^i}}1 {ô}{{\^o}}1  {û}{{\^u}}1
      {Â}{{\^A}}1  {Ê}{{\^E}}1  {Î}{{\^I}}1 {Ô}{{\^O}}1  {Û}{{\^U}}1
      {œ}{{\oe}}1  {Œ}{{\OE}}1  {æ}{{\ae}}1 {Æ}{{\AE}}1  {ß}{{\ss}}1
      {ẞ}{{\SS}}1  {ç}{{\c{c}}}1 {Ç}{{\c{C}}}1 {ø}{{\o}}1  {Ø}{{\O}}1
      {å}{{\aa}}1  {Å}{{\AA}}1  {ã}{{\~a}}1  {õ}{{\~o}}1 {Ã}{{\~A}}1
      {Õ}{{\~O}}1  {ñ}{{\~n}}1  {Ñ}{{\~N}}1  {¿}{{?\`}}1  {¡}{{!\`}}1
      {„}{\quotedblbase}1 {“}{\textquotedblleft}1 {–}{$-$}1
      {°}{{\textdegree}}1 {º}{{\textordmasculine}}1 {ª}{{\textordfeminine}}1
      {£}{{\pounds}}1  {©}{{\copyright}}1  {®}{{\textregistered}}1
      {«}{{\guillemotleft}}1  {»}{{\guillemotright}}1  {Ð}{{\DH}}1  {ð}{{\dh}}1
      {Ý}{{\'Y}}1    {ý}{{\'y}}1    {Þ}{{\TH}}1    {þ}{{\th}}1    {Ă}{{\u{A}}}1
      {ă}{{\u{a}}}1  {Ą}{{\k{A}}}1  {ą}{{\k{a}}}1  {Ć}{{\'C}}1    {ć}{{\'c}}1
      {Č}{{\v{C}}}1  {č}{{\v{c}}}1  {Ď}{{\v{D}}}1  {ď}{{\v{d}}}1  {Đ}{{\DJ}}1
      {đ}{{\dj}}1    {Ė}{{\.{E}}}1  {ė}{{\.{e}}}1  {Ę}{{\k{E}}}1  {ę}{{\k{e}}}1
      {Ě}{{\v{E}}}1  {ě}{{\v{e}}}1  {Ğ}{{\u{G}}}1  {ğ}{{\u{g}}}1  {Ĩ}{{\~I}}1
      {ĩ}{{\~\i}}1   {Į}{{\k{I}}}1  {į}{{\k{i}}}1  {İ}{{\.{I}}}1  {ı}{{\i}}1
      {Ĺ}{{\'L}}1    {ĺ}{{\'l}}1    {Ľ}{{\v{L}}}1  {ľ}{{\v{l}}}1  {Ł}{{\L{}}}1
      {ł}{{\l{}}}1   {Ń}{{\'N}}1    {ń}{{\'n}}1    {Ň}{{\v{N}}}1  {ň}{{\v{n}}}1
      {Ő}{{\H{O}}}1  {ő}{{\H{o}}}1  {Ŕ}{{\'{R}}}1  {ŕ}{{\'{r}}}1  {Ř}{{\v{R}}}1
      {ř}{{\v{r}}}1  {Ś}{{\'S}}1    {ś}{{\'s}}1    {Ş}{{\c{S}}}1  {ş}{{\c{s}}}1
      {Š}{{\v{S}}}1  {š}{{\v{s}}}1  {Ť}{{\v{T}}}1  {ť}{{\v{t}}}1  {Ũ}{{\~U}}1
      {ũ}{{\~u}}1    {Ū}{{\={U}}}1  {ū}{{\={u}}}1  {Ů}{{\r{U}}}1  {ů}{{\r{u}}}1
      {Ű}{{\H{U}}}1  {ű}{{\H{u}}}1  {Ų}{{\k{U}}}1  {ų}{{\k{u}}}1  {Ź}{{\'Z}}1
      {ź}{{\'z}}1    {Ż}{{\.Z}}1    {ż}{{\.z}}1    {Ž}{{\v{Z}}}1  {ž}{{\v{z}}}1
  }
\begin{lstlisting}
@article{gnewuch2025,
  author={Gnewuch, Max Martin and Wahle, Jan Philip and Ruas, Terry and Gipp, Bela},
  title={Big Tech-Funded AI Papers Have Higher Citation Impact, Greater Insularity, and Larger Recency Bias},
  booktitle={2026 International Conference on Artificial Intelligence, Computer, Data Sciences and Applications (ACDSA)},
  year={2026},
  month={02}
}
\end{lstlisting}
\end{tcolorbox}

\vfill
\begin{tikzpicture}
\draw[Primary!40, line width=0.4pt] (0,0) -- (\textwidth,0);
\end{tikzpicture}
\begin{center}
\small\sffamily\textcolor{TextMuted}{Generated \today}
\end{center}

\end{document}